\documentclass[fleqn,usenatbib]{mnras}
    
\usepackage{newtxtext,newtxmath}
\usepackage[T1]{fontenc}
\usepackage{ae,aecompl}

\usepackage{graphicx}	
\usepackage{amsmath}	
\usepackage{amssymb}	
\usepackage{multirow}
\newcommand{\hii}{H$_2\,$}	
\newcommand{\gammah}{$\gamma_{\mathrm{H_2}}$}   
 
\usepackage{color}
\usepackage[normalem]{ulem}

\definecolor{darkgreen}{rgb}{0.13, 0.55, 0.13}

\newcommand{\aref}[1]{\hyperref[#1]{Appendix~\ref{#1}}}
 
\title[H$_2\,$ in Primordial Clouds]{The role of the H$_2$ adiabatic index in the formation of the first stars}

\author[P. Sharda et al.]{
Piyush Sharda,$^{1,2}$\thanks{E-mail: piyush.sharda@anu.edu.au (PS)}
Mark R. Krumholz,$^{1,2}$\thanks{E-mail: mark.krumholz@anu.edu.au (MRK)}
and Christoph Federrath$^{1,2}$
\\
$^{1}$Research School of Astronomy and Astrophysics, Australian National University, Canberra, ACT 2611, Australia\\
$^{2}$ARC Centre of Excellence for All Sky Astrophysics in 3 Dimensions (ASTRO 3D), Australia\\
}
    
\date{Accepted 2019 September 13. Received 2019 September 12; in original form 2019 July 22}
    
\pubyear{2019}

\begin{document}
\label{firstpage}
\pagerange{\pageref{firstpage}--\pageref{lastpage}}
\maketitle
    
\begin{abstract}
The adiabatic index of H$_2\,$ ($\gamma_{\mathrm{H_2}}$) is non-constant at temperatures between $100-10^4\,\mathrm{K}$ due to the large energy spacing between its rotational and vibrational modes. For the formation of the first stars at redshifts 20 and above, this variation can be significant because primordial molecular clouds are in this temperature range due to the absence of efficient cooling by dust and metals. We study the possible importance of variations in $\gamma_{\mathrm{H_2}}$~for the primordial initial mass function by carrying out 80 3D gravito-hydrodynamic simulations of collapsing clouds with different random turbulent velocity fields, half using fixed $\gamma_{\rm H_2} = 7/5$ in the limit of classical diatomic gas (used in earlier works) and half using an accurate quantum mechanical treatment of $\gamma_{\mathrm{H_2}}$. We use the adaptive mesh refinement code FLASH with the primordial chemistry network from KROME for this study. The simulation suite produces almost 400 stars, with masses from $0.02 - 50$ M$_\odot$ (mean mass $\sim 10.5\,\mathrm{M_{\odot}}$ and mean multiplicity fraction $\sim 0.4$). While the results of individual simulations do differ when we change our treatment of $\gamma_{\mathrm{H_2}}$, we find no statistically significant differences in the overall mass or multiplicity distributions of the stars formed in the two sets of runs. We conclude that, at least prior to the onset of radiation feedback, approximating H$_2$ as a classical diatomic gas with $\gamma_{\rm H_2} = 7/5$ does not induce significant errors in simulations of the fragmentation of primordial gas. Nonetheless, we recommend using the accurate formulation of the \hii adiabatic index in primordial star formation studies since it is not computationally more expensive and provides a better treatment of the thermodynamics.
\end{abstract}
    
    
    
\section{Introduction}
\label{s:intro}
Stars are usually classified into three populations based on their metal content \citep{1981ApJ...248..606B,1986MNRAS.223..763M}. The generation of stars with the highest metallicity is known as Population I. Population II corresponds to stars that have relatively less metal content, and Population III is the hypothetical limit of stars that have no metals. Population III stars, also known as first stars, are believed to have formed in dust-free environments out of primordial species produced by the Big Bang \citep{1967Natur.216..976S,1998A&A...335..403G}. They are further classified into Population III.1 (the first generation of stars) and Population III.2 (primordial stars affected by radiation from other stars, see \citealt{2008ApJ...681..771M,2011A&A...533A..32D}). While contemporary star formation is well studied thanks to observations and simulations, the formation of the first generation of stars in the Universe still remains a mystery because of the lack of direct observations at spatially resolved scales beyond $z > 11.1$ \citep{2016ApJ...819..129O}, and of zero-metallicity stars, if any, in the Local Group \citep{2018MNRAS.474..443G,2019MNRAS.482.1204H}. 

The first stars are believed to have formed between redshifts $15 \leq z \leq 30$ (see reviews by \citealt{2002Sci...295...93A,2004ARA&A..42...79B,2005SSRv..117..445G,2005SSRv..116..625C,2013RPPh...76k2901B,2013RvMP...85..809K}), at the center of dark matter minihalos that have high baryonic densities of the order of $\sim 10^4\,\mathrm{cm^{-3}}$ \citep{2002Sci...295...93A,2002ApJ...564...23B}. By this epoch, the first clouds of neutral hydrogen had formed after recombination \citep{1968ApJ...153....1P}. Since the first clouds only contained primordial elements (H, He, Li and their isotopes), cooling during the collapse is inefficient as compared to contemporary star formation where dust and metal lines are present \citep{2005ApJ...626..627O,2013RPPh...76k2901B}. 

Early simulations of the first stars did not have a long dynamical range in time and thus could not follow the large-scale evolution once the primordial clouds started to collapse. They showed no fragmentation, leading to the belief that the first stars were very massive and evolved in isolation \citep{2002ApJ...564...23B,2002Sci...295...93A,2006ApJ...652....6Y}. Once numerical techniques were improved to include modules like sink particles and work with better and more efficient solvers, it became possible to simulate farther in time past the initial collapse. Since then, fragmentation has been observed in almost all simulations of the first stars (for example, \citealt{2011Sci...331.1040C,2012MNRAS.422..290S,2014ApJ...781...60H,2015MNRAS.448..568H}). However, it occurs very close to the central protostar, on scales as small as a few $\mathrm{AU}$ \citep{2006MNRAS.373.1563K,2014ApJ...792...32S,2018arXiv180706248K}. This is because of the lack of an adiabatic core larger than $1\,\mathrm{AU}$ even before protostar formation, as is observed in simulations of contemporary star formation \citep{1969MNRAS.145..271L,1998ApJ...508L..95B}. Thus, in the case of the first stars, the circumstellar disc grows gradually and fragmentation occurs near the central protostar. The observation that primordial gas clouds do fragment naturally raises the question of what initial mass function (IMF) this process yields. Determining the IMF of first stars has thus become a central goal of modern first star research \citep{2004ApJ...612..602T,2006MNRAS.369..825S,2013ApJ...773..185S,2014ApJ...792...32S,2018ApJ...857...46I}.

In this work, we investigate the sensitivity of this IMF, and closely related quantities such as the multiplicity statistics of first stars, to the thermodynamics of molecular hydrogen. This molecule controls the thermal and chemical evolution of collapsing primordial clouds, and becomes the dominant chemical state of hydrogen once the density is high enough. While there has been extensive work on the importance of H$_2$ as a coolant, no published 3D simulations of first star fragmentation to date have systematically investigated another potential role it might play in controlling fragmentation, via the dependence of the adiabatic index on the \hii mass fraction and temperature. 

The adiabatic index is potentially important to the IMF because it determines how easy or hard it is to compress the gas, and thus how much the gas resists fragmentation. A gas with higher $\gamma$ is more resistant to fragmentation because, for the same level of pressure fluctuation, it will respond with a smaller density fluctuation than a gas with lower $\gamma$. In the context of contemporary star formation, \citet{2007ApJ...656L..89B} show that simulations of gravitationally-unstable protoplanetary discs using a correct quantum treatment of \gammah~produce qualitatively different amounts of fragmentation than those where \gammah~is approximated as constant; \citet{2013A&A...550A..52B} show that there are also differences in the subsequent accretion and migration of the fragments. \citet{2014A&A...563A..85V} show that variations in \gammah~lead to changes in the dynamics of the first \citet{1969MNRAS.145..271L} cores that result from collapse. Gravitationally-unstable discs seem particularly sensitive to the adiabatic index of the gas, and this is precisely the mode of fragmentation that determines the IMF of the first stars. Moreover, first star formation occurs in gas clouds at temperatures of hundreds of Kelvin \citep{2005ApJ...626..627O}, which is precisely the temperature range at which the rovibrational modes of \hii first become excited, and thus the departure from classical behaviour is largest. However, no analogous studies have been performed to look for systematic effects of \gammah on formation of the first stars, where at least potentially the effects of variable \gammah~are much larger. The few studies that do include non-constant \gammah~\citep{1983MNRAS.205..705S,1998ApJ...508..141O,2002Sci...295...93A,2006ApJ...652....6Y,2007MNRAS.375..881A,2008ApJ...681..771M,2014MNRAS.444.1566G,2016MNRAS.462.1307S} have not systematically studied its effects, and have also included only variability due to vibrational degrees of freedom, not rotational ones. Our goal in this paper is to carry out a comprehensive study comparing a full quantum mechanical treatment of the \hii molecule to the classical approximation adopted in most earlier 3D simulations.

This paper is organised as follows: \autoref{s:adiabat} discusses how we compute the adiabatic index of H$_{\rm 2}$; \autoref{s:simulation_setup} describes the simulation setup and the physics included; \autoref{s:results} presents our results and findings; finally,  \autoref{s:conclusions} summarises our analysis.

\section{Adiabatic Index of \texorpdfstring{H$_2$}{H2}}
\label{s:adiabat}

The adiabatic index of a gas partly composed of \hii depends on the temperature, mass fraction of \hii and the ratio of ortho to para \hii (which are the two nuclear spin orientations of the molecule, see \citealt{1998ApJ...508..141O,2008MNRAS.388.1627G,doi:10.1063/1.3628453}). To calculate this dependence, we follow the approach of \citet{2014MNRAS.437.1662K}, though equivalent calculations may be found in \citet{2007ApJ...656L..89B} and \citet{2013ApJ...763....6T}. Consider a gas containing multiple chemical species, each with mass fraction $x_{\rm s}$, such that $\sum_{\rm s} x_{\rm s} = 1$. The relation between the net adiabatic index of all species and density is 
    \begin{equation}
    \gamma_{\rm net} = \frac{d\ln P}{d\ln \rho}
    \label{eq:equation1}
    \end{equation}
where $P$ is the pressure. $\rho$ is the volume density, which is related to the number density ($n_{\mathrm{s}}$) and mass fraction ($x_{\mathrm{s}}$) as
    \begin{equation}
    n_{\mathrm{s}} = \frac{x_{\mathrm{s}}\rho}{A_{\mathrm{s}}m}\,
    \end{equation}
where $m$ is one a.m.u., and $A_{\mathrm{s}}$ is the mass number of the species. The net adiabatic index for the system can be written as the ratio of specific heats at constant pressure and volume
    \begin{equation}
    \gamma_{\mathrm{net}} =  \frac{c_p/k_{\mathrm{B}}}{c_v/k_{\mathrm{B}}} = \frac{c_v/k_{\mathrm{B}} + 1}{c_v/k_{\mathrm{B}}},
    \end{equation}
where $c_p$ and $c_v$ are the specific heats per H nucleon at constant volume and pressure, respectively. We obtain these from the internal energy per unit volume,
    \begin{equation}
    e_{\mathrm{g}} = n_{\mathrm{H}} k_{\mathrm{B}} T \frac{\mathrm{d\,ln\,}z}{\mathrm{d\,ln\,}T}\,,
    \end{equation}
where $z$ is the ensemble partition function given by the product of partition functions for the translational, rotational and vibrational degrees of freedom $z = Z_{\mathrm{trans}}Z_{\mathrm{rot}}Z_{\mathrm{vib}}$, $T$ is the temperature and $n_{\rm H}$ is the number density of H nuclei (which is invariant under chemical reactions). The specific heat per H nucleon at constant volume is related to $e_{\rm g}$ by
    \begin{equation}
    \frac{c_{v}}{k_B} = \frac{1}{n_{\rm H}} \frac{\partial e_{\mathrm{g}}}{\partial T}\,.
    \end{equation}
Using partition functions and mass fractions for ortho and para H$_2$ ($Z_{\mathrm{rot}} = Z_{\mathrm{pH_2}}\,Z_{\mathrm{oH_2}}$, as defined below), this becomes:
    \begin{equation}
    \begin{split}
    \frac{c_{v}}{k_{\mathrm{B}}} = \frac{3}{2} + x_{\mathrm{pH_2}}\,\frac{\partial}{\partial T} \bigg(\frac{T^2}{Z_{\mathrm{pH_2}}} \frac{\partial Z_{\mathrm{pH_2}}}{\partial T}\bigg) + x_{\mathrm{oH_2}}\,\frac{\partial}{\partial T} \bigg(\frac{T^2}{Z_{\mathrm{oH_2}}} \frac{\partial Z_{\mathrm{oH_2}}}{\partial T}\bigg)\\
    + (x_{\mathrm{oH_2}} + x_{\mathrm{pH_2}}) \frac{\theta^2_{\mathrm{vib}}\,\mathrm{exp}(-\theta_{\mathrm{vib}}/T)}{T^2 [1 - \mathrm{exp}(-\theta_{\mathrm{vib}}/T)]^2}
    \end{split}
    \label{eq:5}
    \end{equation}
where $x_{\mathrm{H_2}} = x_{\mathrm{oH_2}} + x_{\mathrm{pH_2}}$, and we have assumed that all species other than H$_2$ have no internal degrees of freedom. While an exact calculation of the partition function should also include contributions from electronic and nuclear degrees of freedom, these modes are not excited in the range of temperatures relevant to this study; hence they can safely be ignored, and we can simply adopt $\gamma = 5/3$ for monoatomic species like He. Similarly, we ignore the effects of overlap between higher vibrational levels, vibrational continuum and electronically excited levels of \hii that occur at temperatures much higher than those we study in this work. We also use a fixed ortho:para ratio for reasons we discuss further below. The last term in \autoref{eq:5} corresponds to vibrational degrees of freedom of H$_2$, where $\theta_{\mathrm{vib}} = 5987$ K \citep{1983ApJ...264..485D}.
    
The rotational partition functions of para- and ortho-H$_2$ are given by
\begin{eqnarray}
Z_{\mathrm{pH_2}} & = & \sum_{J\,\mathrm{even}} (2J+1)\,\mathrm{exp}\,\bigg[-\frac{J(J+1)\theta_{\mathrm{rot}}}{T}\bigg] \\
Z_{\mathrm{oH_2}} & = & e^{2\theta_{\rm rot}/T} \bigg(\sum_{J\,\mathrm{odd}} 3(2J+1)\,\mathrm{exp}\,\bigg[-\frac{J(J+1)\theta_{\mathrm{rot}}}{T}\bigg]\bigg)
\end{eqnarray}
where $\theta_{\mathrm{rot}} = 85.4$ K \citep{1975ApJ...199..619B}. The leading exponential term in the ortho \hii partition functions ensures that rotation only contributes to internal energy when the rotational states are excited \citep{2007ApJ...656L..89B}.

\begin{figure}
\includegraphics[width=\columnwidth]{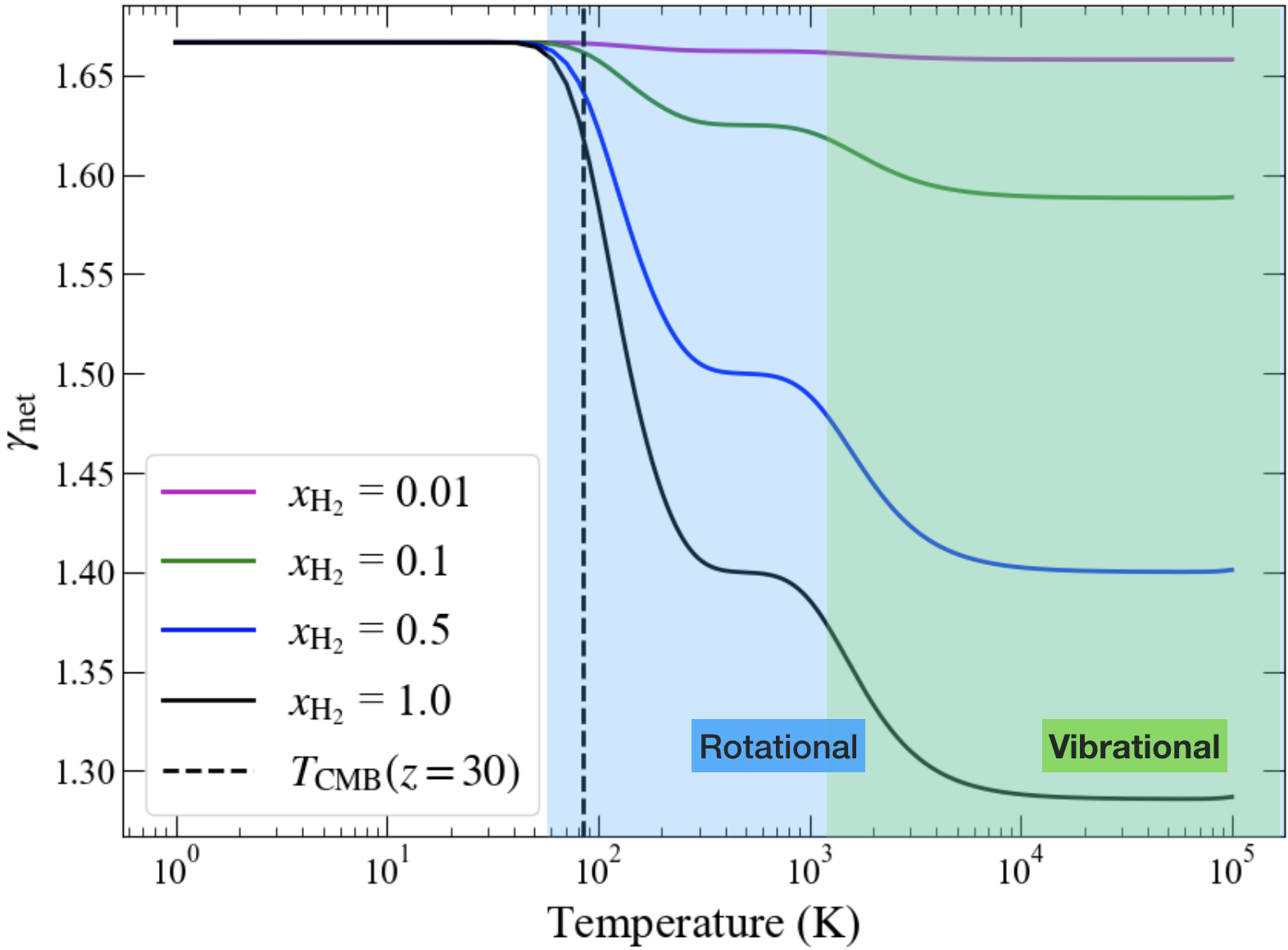}
\caption{Net adiabatic index ($\gamma_{\mathrm{net}}$) as a function of temperature for primordial gas with varying fractional abundances of \hii, assuming an ortho- to para-ratio of 3:1 and no other species have any internal degrees of freedom. The dashed-black line marks the cosmic microwave background (CMB) floor at $z=30$. The temperature range indicated in blue is dominated by the rotational degrees of freedom of \hii whereas that in green is dominated by its vibrational degrees of freedom. The deviation of $\gamma_{\mathrm{net}}$ from the standard values is greatest for a completely molecular gas, and negligible if $x_{\mathrm{H_2}} \lesssim 0.01$.}
\label{fig:gammaplot}
\end{figure}

\autoref{fig:gammaplot} shows the variation of the net adiabatic index of the system ($\gamma_{\mathrm{net}}$) as a function of temperature ($T$) at different mass fractions of H$_2$ ($x_{\mathrm{H_2}}$), assuming an ortho- to para-ratio of 3:1 (see below). When the gas is completely molecular (\textit{i.e.,} \gammah = $\gamma_{\mathrm{net}}$), it can be described as monoatomic (3 translational degrees of freedom) at low temperatures ($T < 50\,\mathrm{K}$) with \gammah = 5/3, and diatomic at high temperatures (3 translational, 2 rotational and 2 vibrational degrees of freedom) where the continuum limit is reached ($T \gtrsim 10^4\,\mathrm{K}$) with \gammah = 9/7.\footnote{Since H$_2$ is collisionally dissociated at temperatures well below 10000 K, in reality it never reaches the high temperature continuum limit.} Primordial star formation sits squarely in between these two regimes, where first the rotational modes are excited during collapse and then the vibrational modes are excited in accretion shocks around first stars, leading to the complex behaviour of \gammah as a function of $T$ shown in \autoref{fig:gammaplot}. 

It should be noted that our calculation of the adiabatic index depends on our choice of the ratio of ortho-H$_2$ to para-H$_2$, and any possible dependence of this ratio on temperature or density. However, \cite{2008MNRAS.388.1627G} show that the ortho-to-para is not very sensitive to temperature at the redshifts important for Population III star formation, and the standard assumption of an ortho-to-para ratio of 3:1, \textit{i.e,} $x_{\rm oH_2} = 3 x_{\rm pH_2}$, as usually found in the present-day Universe \citep{1999ApJ...516..371S}, produces results similar to a more detailed treatment. Due to interconversions facilitated by collisions with H$^+$ in the primordial gas, this ratio drops down to 0.25:1 at $z \approx 20$ in environments where the mass fraction of \hii drops to $10^{-6}$ \citep{2000MNRAS.316..901F,2007MNRAS.377..705F}, but at such low H$_2$ abundances, the value of $\gamma_{\mathrm{net}}$ is essentially independent of \gammah in any event (\autoref{fig:gammaplot}). Keeping these studies in mind, we fix the ortho-to-para ratio to be 3:1 for our simulations. 

\section{Numerical and Physical Ingredients}
\label{s:simulation_setup}

\subsection{Numerical Hydrodynamics}
\label{s:hydro}
We use the adpative mesh refinement (AMR, \citealt{1989JCoPh..82...64B}) code FLASH \citep{2000ApJS..131..273F,2008ASPC..385..145D}. We utilize an approximate Riemann solver for our hydrodynamic simulations \citep{Bouchut2007,Bouchut2010} which was developed for FLASH by \citet{2009JCoPh.228.8609W} and \citet{2011JCoPh.230.3331W}. We treat the self-gravity of the gas with a tree-based solver \citep{2018MNRAS.475.3393W}. We use the sink particle technique developed for FLASH \citep{2010ApJ...713..269F,2011IAUS..270..425F,2014ApJ...797L..19F} to follow the evolution of collapsing gas at high resolutions at late times. Sink particles are frequently used in hydrodynamic simulations of star formation as a proxy for stellar sources \citep{1995MNRAS.277..362B,2004ApJ...611..399K,2005A&A...435..611J,2010ApJ...709...27W,2011ApJ...730...40P,2013ApJS..204....8G,2013MNRAS.430.3261H,2014MNRAS.445.4015B,2018MNRAS.480.2562J}. These Lagrangian particles can travel inside the grid, accrete gas and contribute to the gravitational potential in the region. The sink particle method developed in \cite{2010ApJ...713..269F} uses a rigorous set of checks to ensure that only Jeans unstable gas that is converging and bound and has a gravitational potential minimum in cells at the sink density threshold at the highest level of refinement is converted into a sink, thus avoiding artificial fragmentation. The density threshold where sink particles are created at the standard resolution used in our simulations (see below) is $1.45\times10^{-11}\,\mathrm{g\,cm^{-3}}$. We use the distribution of sink masses to study fragmentation around the primary sink. As the numerical scale we use in this work is much larger than the radii of actual protostars, we do not allow the merging of sink particles in our simulations (see, for example, \citealt{2014ApJ...792...32S,2018MNRAS.479..667R}).

In order to completely encompass the cloud that collapses to form stars, we define a cubical box of size $L = 2.4\,\mathrm{pc}$ to run our simulations. We set the boundary conditions to be outflow-type to enable mass-loss from the cloud, if any, during star formation\footnote{Note that outflow in FLASH also means that inflow can occur.}. The boundary condition for gravity is `isolated' (\textit{i.e.}, not periodic). We use a base grid of $8^3$ cells plus 14 levels of refinement in this work, which results in a unit cell length at the highest level of refinement of $dx = 7.6\,\mathrm{AU}$ and a maximum resolvable density $n \sim 10^{15}\,\mathrm{cm^{-3}}$; the maximum effective resolution of the simulation is $65536^3$. This choice is motivated by optimizing the trade-off between higher resolution and computational costs. We repeat three representative simulations with different random seeds of turbulence (see \autoref{s:init_conds}) at 12, 13, 14 and 15 levels of refinement to check numerical convergence; we present the results of our convergence study in \aref{s:append_convergence} and show that it is reasonable to believe convergence has been achieved to first order at resolution 14.

Numerous studies have shown that it is important to resolve the scales at which turbulence can amplify magnetic fields through small-scale dynamo action \citep{2010ApJ...721L.134S,2010ApJ...713..269F,2010A&A...522A.115S,2012ApJ...754...99S,2012SSRv..169..123B,2012PhRvE..85b6303S,2013NJPh...15a3055B,2013MNRAS.432..668L}. The required resolution in this case is at least 30 cells per Jeans length \citep{2011ApJ...731...62F}, which is 7.5 times more than the Truelove criterion to avoid artificial collapse in gravito-hydrodynamic simulations \citep{1997ApJ...489L.179T}. Although we do not include magnetic fields in this work, we satisfy the criterion suggested by \cite{2011ApJ...731...62F} by using 32 cells per Jeans length, to maintain self-consistency with other works (P. Sharda et al., in prep.). In fact, using less than 30 grid cells per Jeans length leads to underestimates not only of the amplification of magnetic fields, but also of the amount of kinetic energy that is resolved on the Jeans scale \citep{2011ApJ...731...62F} and the structure of the gas (for example, the scale height of accretion discs; see \citealt{2014ApJ...797L..19F}).

\subsection{Primordial Chemistry}
\label{s:krome}
We utilize the KROME package for primordial chemistry, which has been developed to include chemistry in hydrodynamic simulations for astrophysical applications \citep{2014MNRAS.439.2386G}. KROME uses a subroutine of pre-designed and re-writable chemical networks for various astrophysical phenomena which can be embedded in numerical codes like FLASH. It uses the differential solver DLSODES \citep{Hindmarsh:1980:LLT:1218052.1218054,osti_15013302} to solve the reaction network and evolves the temperature and density of the system in accordance with the chemistry and the specified heating and cooling processes \citep{2013MNRAS.431.1659G,2013MNRAS.434L..36B}. The network of primordial chemical reactions we use in our simulations is \texttt{react\_primordial\_3} which is the most robust primodial chemistry network and includes the following species: H, H$_2$, H$^+$, H$^-$, He, He$^+$, He$^{++}$, H$^{+}_{2}$ and $\mathrm{e^-}$. 

We include a variety of chemical and radiative heating and cooling processes, all of which are computed by KROME. The cooling processes we include are: 1) cooling by \hii through excitation of rovibrational modes in \hii and subsequent emission of photons, 2) cooling through collisionally induced emission (CIE) which occurs due to the formation of `supermolecules' with finite electric dipole from collisions between different molecules, 3) cooling due to endothermic chemical reactions, 4) atomic cooling due to collisional ionisation, collisional excitation and recombination of primordial species and bremsstrahlung emission from ionised species, and 5) cooling due to Compton scattering of cosmic microwave photons by free electrons. In addition, we impose a constraint on the minimum temperature such that it never decreases beyond the cosmic microwave background temperature at the assumed redshift ($T_{\mathrm{CMB}}(z=30) = 84.63\,\mathrm{K}$). The heating processes we include are 1) chemical heating generated from reaction enthalpies and 2) compressional heating (as computed by the hydrodynamics module). 

At high densities, the cooling rates are suppressed by opacity effects. For cooling due to \hii, we approximately account for this by using the H$_2$ cooling function provided by \cite{2004MNRAS.348.1019R}. This approximation diverges from the more detailed treatment of opacity by \citealt{2013ApJ...763...52H} (see also, \citealt{2014MNRAS.444.1566G,2015ApJ...799..114H}) when $x_{\mathrm{H_2}}\lesssim\,0.5$. However, in practice these cooling functions differ only where the gas is dense enough to be optically thick, and has also been heated by adiabatic compression to the point where H$_2$ undergoes significant collisional dissociation. Such conditions prevail only at densities $\gtrsim 10^{16}$ cm$^{-3}$, an order of magnitude higher than those we resolve. Thus, over the density range we cover, the \citet{2004MNRAS.348.1019R} and \citet{2013ApJ...763...52H} \hii cooling functions are very similar. Apart from this, the Lyman-$\alpha$ cooling formulation that we include in our simulations (in KROME) diverges from its true value in optically-thick regimes where both the densities and temperatures are high; such regions constitute the accretion disks around sink particles, as we later show in Section \ref{s:results}. However, we do not expect this effect to significantly alter the temperature because the Lyman-$\alpha$ cooling rate is extraordinarily sensitive to temperature and only very weakly sensitive to optical depth (see, for example, Section 2.1 of \citealt{2017MNRAS.472.2773G}); thus even fairly large optical depths alter the temperature relatively little. We also omit cooling due to H$^-$. While this can be important in regions where H$^-$ is abundant, due to its large cross section, the H$^-$ abundance is very low at densities $\lesssim 10^{15}$ cm$^{-3}$ \citep{2001ApJ...546..635O,2014A&A...572A..22V}, the highest we resolve in this work.

Our chosen chemical network does not include deuterium, which was also produced by the Big Bang \citep{1976Natur.263..198E}. We choose to omit it because deuterium has no significant impact on the adiabatic index because of the low fractional abundance of HD as compared to H or \hii. HD can be an important coolant in low density regions ($10^5 \leq n \leq 10^8\,\mathrm{cm^{-3}}$) at temperatures of the order of $100\,\mathrm{K}$ \citep{2002P&SS...50.1197G,2005MNRAS.364.1378N,2007ApJ...663..687Y} in cases where the primordial gas does not go through an ionized phase \citep{2006MNRAS.366..247J,2008MNRAS.388.1627G}. However, as we show below, our simulations start at $n \sim 10^4\,\mathrm{cm^{-3}}$ and fragmentation occurs at densities that are $10^{4-5}$ times the density range quoted above. Similarly, we do not include primordial Li \citep{2011ARNPS..61...47F}, since it has been shown that its contribution to both chemistry and cooling is unimportant \citep{1984ApJ...280..465L,Lepp_2002,2013ARA&A..51..163G,2018MNRAS.476.1826L}.

\subsection{Initial Conditions}
\label{s:init_conds}
We initiate our simulations by setting up a spherical cloud core with a homogeneous density. Taking inspiration from cosmological simulations that form dark matter minihalos where baryonic cores form in overdense regions, we begin from a core of mass $M_{\mathrm{core}} = 1000\,\mathrm{M_{\odot}}$ and radius $R_{\mathrm{core}} = 1\,\mathrm{pc}$ \citep{2000ApJ...540...39A,2002Sci...295...93A,2002ApJ...564...23B}. These parameters are similar to that for Bonnor-Ebert spheres on the verge of collapse, and are often used in such simulations as initial conditions (for example, \citealt{2013MNRAS.435.3283M,2014ApJ...792...32S,2014ApJ...781...60H,2014ApJ...785...73S,2016MNRAS.460.2432H,2018MNRAS.479..667R}). Our initial density ($n_{\mathrm{core}} = 9050\,\mathrm{cm^{-3}}$) is thus in good agreement with the overdensity observed in cosmological simulations. Based on 1D calculations of primordial cloud collapse using KROME that we run from low densities ($n = 1\,\mathrm{cm^{-3}}$) and temperatures ($T = 100\,\mathrm{K}$), we find that the temperature reaches 265 K by the time the density has reached $n \sim 10^4\,\mathrm{cm^{-3}}$. Thus, we set $T_{\mathrm{core}} = 265\,\mathrm{K}$. This 1D model also sets the initial mass fractions of all species for our simulations. Specifically, we use $x_{\mathrm{H}} = 0.7502,\,x_{\mathrm{H_2}} = 0.0006$ and $x_{\mathrm{He}} = 0.2492$, which also agree well with initial mass fractions for several other simulations at the same initial temperature and density; the He abundance is that predicted by the Big Bang nucleosynthesis \citep{2007ARNPS..57..463S,2013ARA&A..51..163G}. To ensure the simulation box is in pressure equilibrium, we set the corresponding background density and temperature to be 100 times lower and higher, respectively. We put the initial core into solid body rotation around the $\hat{z}$ axis, with the initial angular velocity set such that the rotational energy is 3 per cent of the gravitational energy. This choice is motivated by the angular momentum of minihalos observed in cosmological simulations \citep{2002ApJ...564...23B,2006ApJ...652....6Y}, and is roughly what is expected for a random turbulent field \citep{1993ApJ...406..528G,2000ApJ...543..822B,2018MNRAS.477.4241L}.

\begin{table}
\centering
\caption{Initial conditions of the spherically homogeneous primordial cloud.}
\label{tab:inicond}
\begin{tabular}{|lcr|}
\hline
Parameter & Symbol & Value\\
\hline
Cloud Mass & $M_{\mathrm{core}}$ & $1000\,\mathrm{M_{\odot}}$\\
Cloud Radius & $R_{\mathrm{core}}$ & $1\,\mathrm{pc}$\\
Cloud Number Density & $n_{\mathrm{core}}$ & $9050\,\mathrm{cm^{-3}}$\\
Cloud Temperature & $T_{\mathrm{core}}$ & $ 265\,\mathrm{K}$\\
Rotational / Gravitational Energy & $E_{\mathrm{rot}}/E_{\mathrm{grav}}$ & $0.03$\\
Mass Fraction of H & $x_{\mathrm{H}}$ & $0.7502$\\
Mass Fraction of \hii & $x_{\mathrm{H_2}}$ & $0.0006$\\
Mass Fraction of He & $x_{\mathrm{He}}$ & $0.2492$\\
CMB Temperature at $z=30$ & $T_{\mathrm{CMB}}$ & $84.63\,\mathrm{K}$\\
Turbulence & $v_{\mathrm{rms}}$ & $1.84\,\mathrm{km\,s^{-1}}$\\
Sound Speed & $c_{\mathrm{s}}$ & $1.84\,\mathrm{km\,s^{-1}}$\\
\hline
\end{tabular}
\end{table}

Our initial velocity includes a random turbulent component on top of the organised rotational field. We only change the random seed value of turbulence between different runs. Our reasons for including turbulence are two-fold: 1) cosmological simulations show that turbulence is driven in dark matter minihalos by the motion of baryons towards the center of the minihalo, leading to the formation of overdense regions ($n \sim 10^4\,\mathrm{cm^{-3}}$) where collapse takes place \citep{2008MNRAS.387.1021G,2012MNRAS.419.3092P,2018A&A...610A..75C}; and 2) turbulence can also be generated by streaming velocities between the dark matter and baryons \citep{2014Natur.506..197F} or primordial magnetic fields \citep{1996PhRvD..54.1291B,2016PhyS...91j4008K}. Taking this into account and following \cite{2008MNRAS.387.1021G}, we introduce rms velocity fluctuations ($v_{\mathrm{rms}}$) equal to the sound speed ($c_{\mathrm{s}}$) in the simulation box (\textit{i.e.,} we set an initially sonic turbulence with Mach 1; see also, \citealt{2011ApJ...727..110C,2012ApJ...754...99S,2018MNRAS.479..667R}). The initial turbulent velocity field that we add has a power spectrum $P_{\mathrm{v}} \sim k^{-1.8}$ from wavenumbers $k/(2\pi/L) = 2-20$ where $L$ is the side length of the computational domain. We choose the above scaling to model sonic turbulence that we include, which lies between the Kolmogorov turbulence ($k^{-1.67}$, for incompressible subsonic fluids) and Burgers turbulence ($k^{-2}$, for compressible supsersonic fluids) and has been studied in detail in numerous works (for example, \citealt{2007ApJ...665..416K,2013MNRAS.436.1245F}). We summarise all properties of our initial conditions in \autoref{tab:inicond}.

\section{Results}
\label{s:results}

We carry out two sets of simulations. One set uses a fixed value $\gamma_{\rm H_2}=7/5$, as is the common practice in first stars simulations; we refer to these runs as \textit{Fixed \gammah}. The second set uses a value of \gammah~computed via a full quantum mechanical treatment, as described in \autoref{s:adiabat}; we refer to these as the \textit{Variable \gammah} simulations. We carry out 40 realisations of each type of simulation, using different turbulent velocity fields. Velocity fields are matched in pairs of fixed and variable \gammah~simulations, so the same 40 turbulent fields are used in each simulation set. We note that simulations with variable \gammah are not computationally expensive, and take the same time and resources as those with fixed \gammah. Thus, irrespective of the results, we highly recommend variable \gammah formulation be used for future studies of primordial star formation since it is more accurate. We define a sink formation efficiency 
\begin{equation}
    \mathrm{SFE} = \frac{\sum M_{\mathrm{sink}}}{M_{\mathrm{core}}}\,
    \label{eq:eqSFE}
\end{equation}
and present all analysis at $\mathrm{SFE} = 5\%$. In other words, the analysis and figures we present is at the time when the sink particles have collectively accreted 5 per cent of the initial cloud mass. The reason for this is radiation feedback, which is not included in our simulations, can inhibit the growth of massive protostars $\gtrsim 25\,\mathrm{M_{\odot}}$ \citep{2011Sci...334.1250H,2015MNRAS.454.2441S,2016ApJ...824..119H}. By limiting our analysis to the time when a relatively small mass has been accreted, we limit ourselves to considering the time before which our simulations will substantially deviate from reality.

\subsection{Qualitative Outcome}
\label{s:phys_props}
\begin{figure*}
\includegraphics[width=0.95\textwidth]{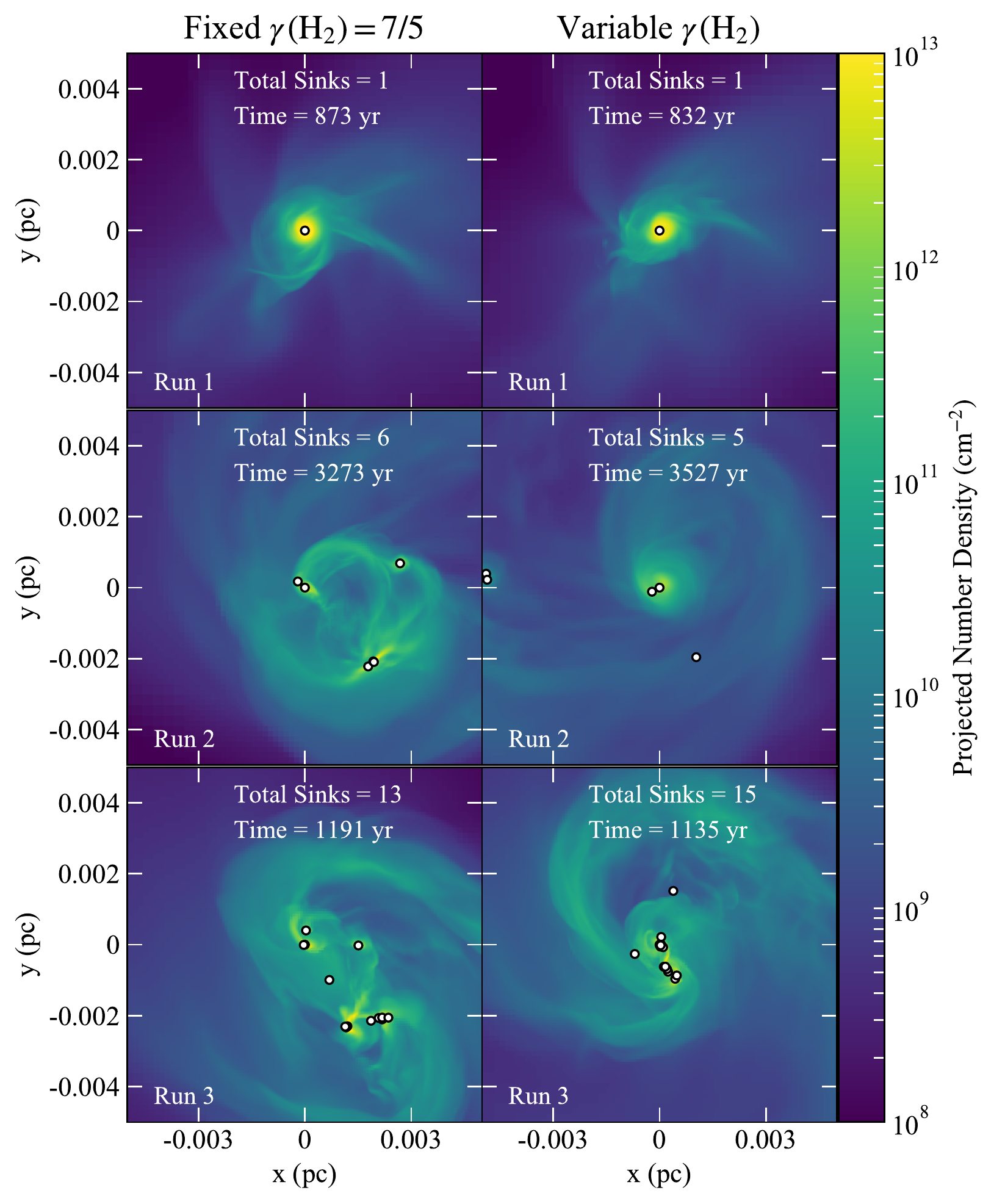}
\caption{Face-on density-weighted projection maps of the number density (through the $\hat{z}$ axis) for a pair of three representative runs showing no (\textit{top panels}), some (\textit{middle panels}) and high fragmentation (\textit{bottom panels}) for fixed (left) and variable (right) \gammah, respectively. All the snapshots are taken when the sink(s) (shown in white circles with black boundaries) have collectively accreted 5 per cent of the initial cloud mass ($\mathrm{SFE} = 5\%$, see \autoref{eq:eqSFE}). The snapshots cover a spherical region of radius $0.01\,\mathrm{pc}$, centered on the most massive sink in the simulation. The time printed in each panel is the time since the formation of the first sink particle in each run. Each of the paired fixed and variable \gammah cases shown begins from identical initial conditions, so the differences seen in the corresponding maps are solely due to variations in \gammah.}
\label{fig:snaps_numdens}
\end{figure*}

\begin{figure*}
\centering
\includegraphics[width=0.95\textwidth]{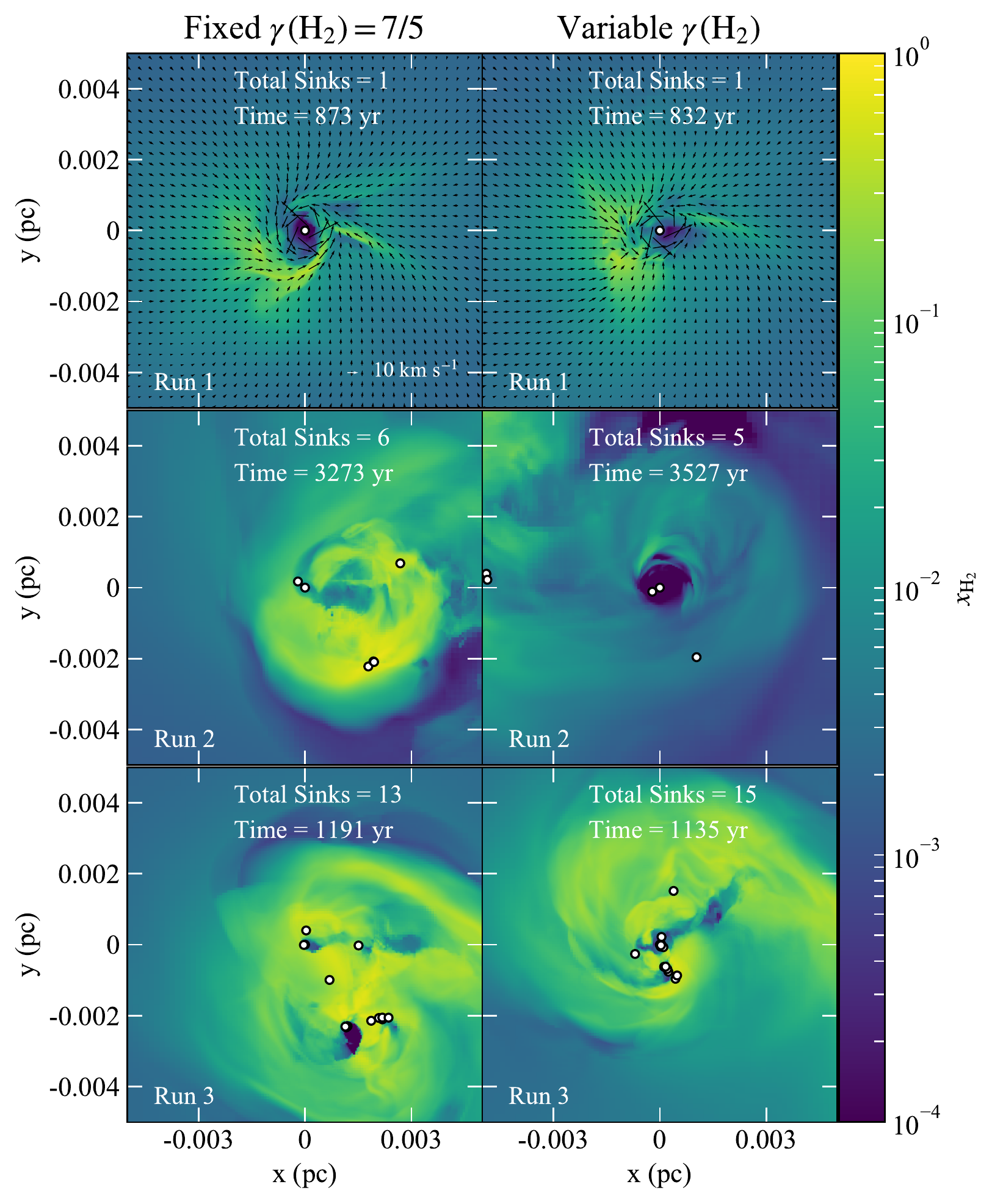}
\caption{Same as  \autoref{fig:snaps_numdens} but showing the density-weighted mean mass fraction of \hii ($x_{\mathrm{H_2}}$). Quivers plotted on the top panels represent the velocity vectors.}
\label{fig:snaps_h2}
\end{figure*}

\begin{figure*}
\includegraphics[width=0.95\textwidth]{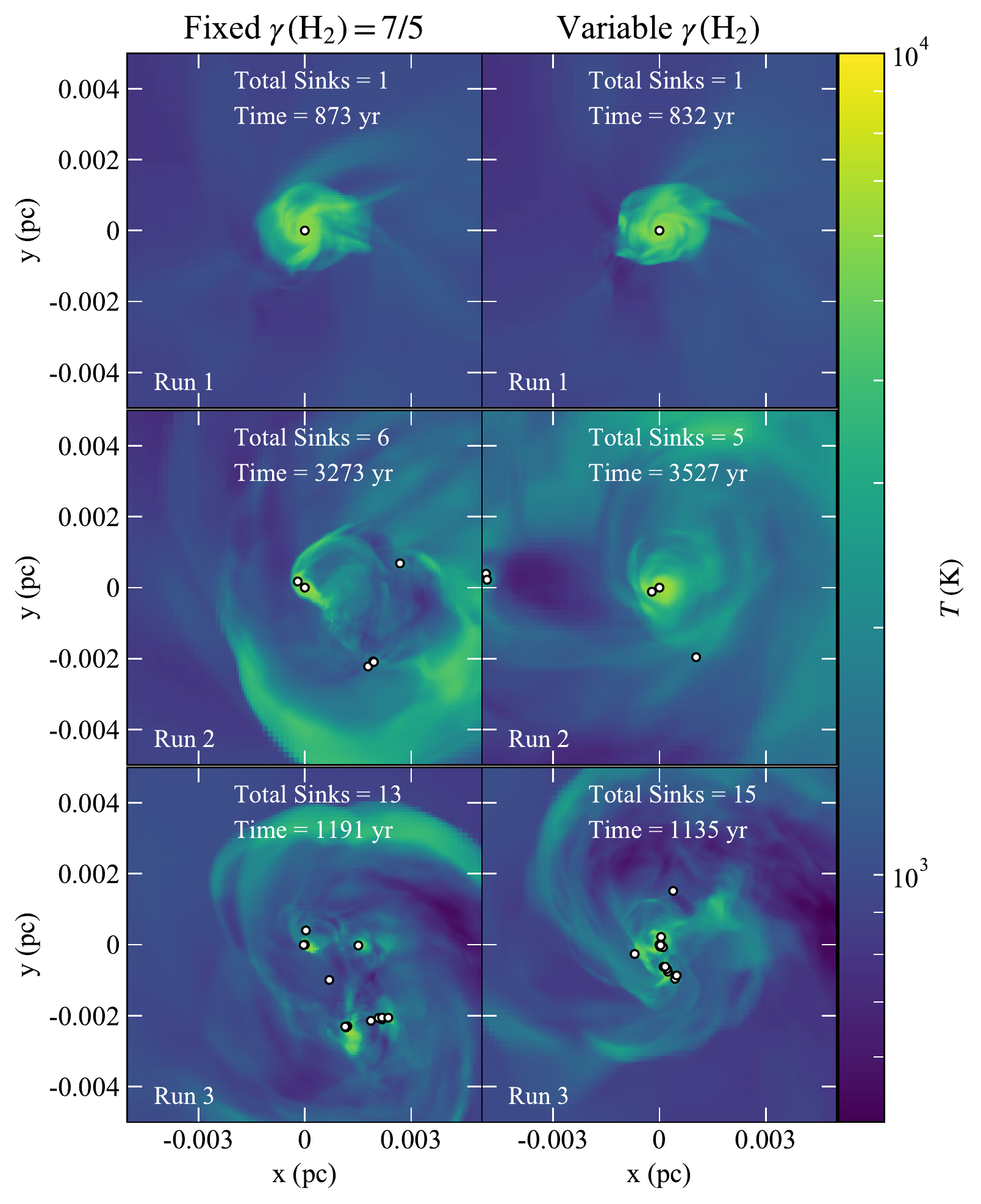}
\caption{Same as  \autoref{fig:snaps_numdens} but showing the density-weighted mean temperature.}
\label{fig:snaps_temp}
\end{figure*}

\autoref{fig:snaps_numdens} shows the density-weighted projections of number density (through the $\hat{z}$ axis) in three representative runs from the fixed and variable \gammah cases where we find no (top panel), some (middle panel) and high fragmentation (bottom panel), respectively\footnote{\label{footnote:3}A movie showing the evolution of density and \gammah as collapse and fragmentation occur in a representative run is available as supplementary online material.}. The white dots marked with black boundaries in each panel represent the locations of sink particles in the corresponding runs. All the projections are focused on the $0.01\,\mathrm{pc}$ region around the most massive sink particle. While the runs with no fragmentation after the first sink is formed show a dense accretion disk around it, we detect diverse filamentary and spiral structures around the sink particles in other runs where subsequent fragmentation has occurred. The densities we reach in the simulations are of the order of $10^{15}\,\mathrm{cm^{-3}}$ where we begin to run into optically thick media. However, we expect the optically thick cooling rate approximation used from \cite{2004MNRAS.348.1019R} in KROME to remain accurate, because the densest regions that we resolve are almost fully molecular, as we notice from \autoref{fig:snaps_h2}, which shows the mass fraction of \hii in the same region as illustrated in \autoref{fig:snaps_numdens}. However, where strong accretion shocks are present, \hii has been dissociated into H. The presence of shocks can be seen through the velocity quivers overplotted on the pair of projection maps in the upper panel of \autoref{fig:snaps_h2} and the temperature field shown in \autoref{fig:snaps_temp}. The temperature range has a strict lower-limit given by the CMB temperature at our assumed redshift ($z=30$, see \autoref{tab:inicond}) as well as a loose upper-limit set by the onset of atomic cooling at temperatures greater than $10^4\,\mathrm{K}$. The typical Mach numbers we find in the runs are between $2-35$.

The qualitative outcome of our simulations, including the diversity in level of fragmentation are similar to the results of other simulations of first star formation that include turbulence \citep{2012ApJ...745..154T,2012ApJ...754...99S,2018MNRAS.479..667R}. Stars forming in highly-fragmented systems often experience fragmentation-induced starvation that limits the gas per star available for accretion \citep{2006MNRAS.373.1563K,2010ApJ...725..134P,2012MNRAS.420..613G}. This effect is more prominent for Population III star formation than for contemporary star formation, due to the smaller distances from the the primary at which fragmentation occurs.

\begin{figure*}
\includegraphics[width=1.0\columnwidth]{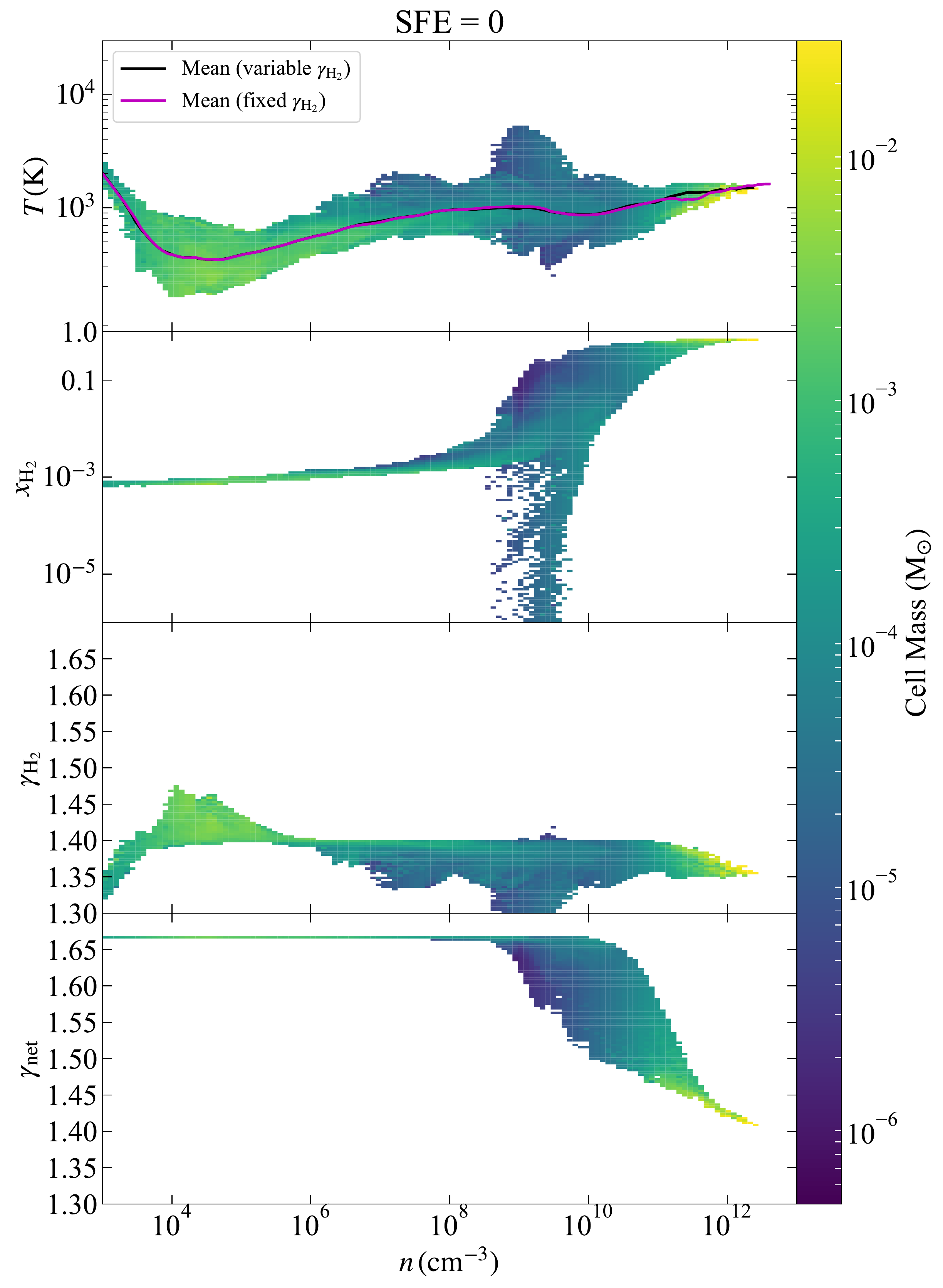}
\includegraphics[width=1.0\columnwidth]{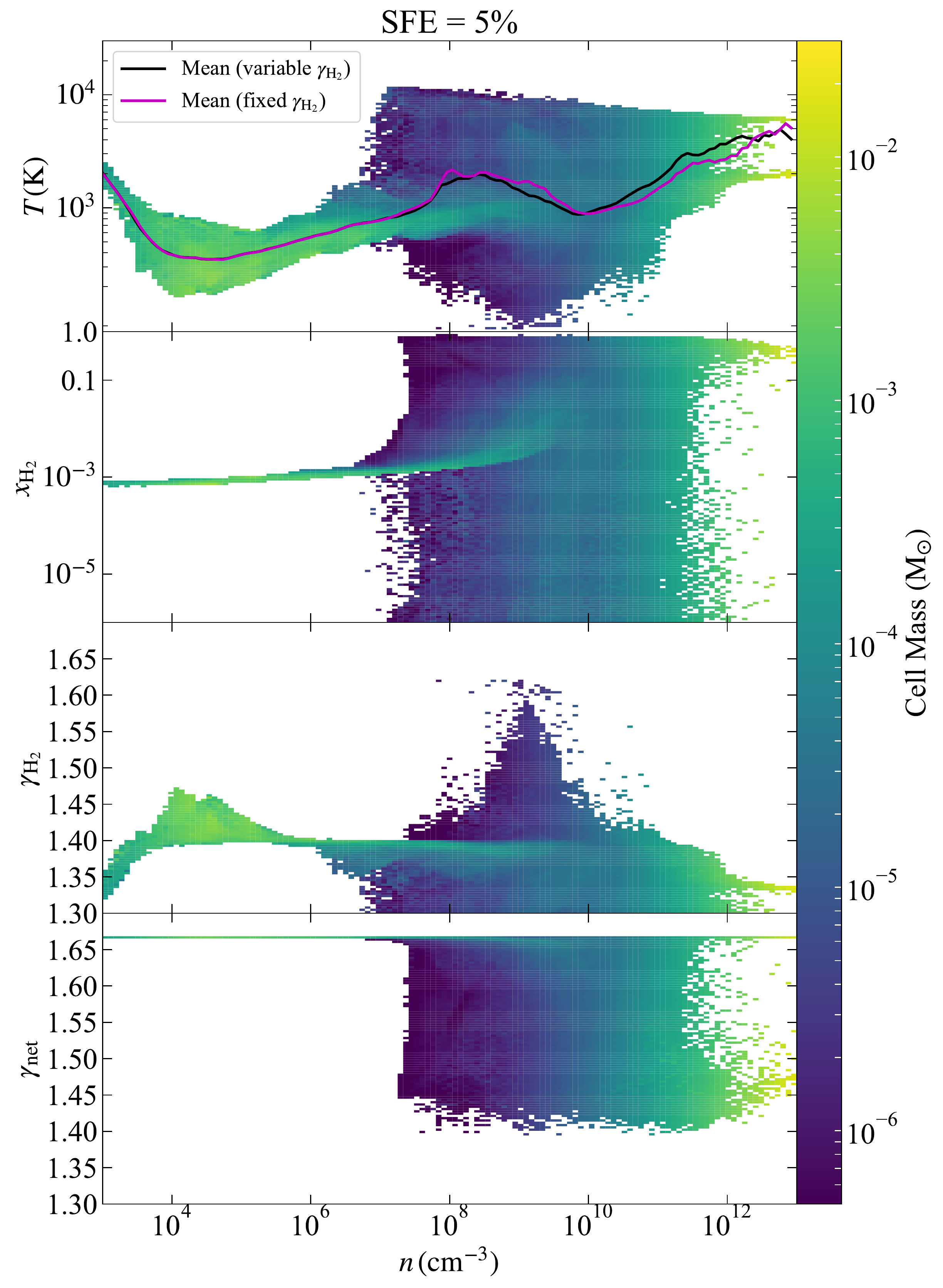}
\caption{Joint distributions of number density ($n$) as a function of temperature ($T$, \textit{first panel}), mass fraction of \hii ($x_{\mathrm{H_2}}$, \textit{second panel}), adiabatic index of \hii alone (\gammah, \textit{third panel}) and of all gas ($\gamma_{\rm net}$, \textit{fourth panel}) in a spherical volume of radius $0.5\,\mathrm{pc}$ centered on the most massive sink particle. The two figures reflect the characteristics of the system just before the formation of the first sink and at $\mathrm{SFE}=5\%$. They belong to one of the 40 runs randomly selected from the variable \gammah subset. Also plotted in the top panels is the mean trend of $T$ as a function of $n$ for the variable (black) and the corresponding fixed \gammah (magenta) run.}
\label{fig:phaseplots}
\end{figure*}

\autoref{fig:phaseplots} shows the joint distributions of number density as a function of temperature, mass fraction of \hii ($x_{\mathrm{H_2}}$), adiabatic index of \hii (\gammah) and the net adiabatic index of all species ($\gamma_{\mathrm{net}}$) for a representative simulation of the variable \gammah case. We show these characteristics just before the formation of the first sink particle and at the end point of our simulations where $\mathrm{SFE}=5\%$. We sample these distributions over all the cells within $0.5\,\mathrm{pc}$ of the most massive sink in the simulation at $\mathrm{SFE} = 5\%$. The evolution of temperature with density in the collapsing cloud closely follows the one zone model of \cite{2005ApJ...626..627O}, as can be noticed from the mean value of the $n-T$ distribution plotted as the black curve on the top panels in Figure \ref{fig:phaseplots}. For comparison, we also plot the mean value of the corresponding fixed \gammah case in magenta. It is clear that the mean value between the two cases only slightly differs throughout the collapse of the cloud. The diverging behavior from the mean at higher densities is due to the formation of accretion discs around sink particles that contain a huge diversity of cells with different positions in the $n-T$ space. There is a clear scatter in the distributions that is a result of variations in temperature and mass fraction of \hii, and the variance of the distribution of \gammah increases monotonically with time. The distributions have a number of features whose physical origin is easy to understand. At densities $n\lesssim 10^7$ cm$^{-3}$, $\gamma_{\rm net}$ is very close to $5/3$ because the H$_2$ fraction is tiny, as can be noticed from the second panel of \autoref{fig:phaseplots}. Only above this density does an appreciable \hii fraction build up due to 3-body reactions \citep{2005ApJ...626..627O,2008MNRAS.388.1627G,2013MNRAS.431.1659G}; it also undergoes rapid dissociation due to high temperatures, thus leading to negligible $x_{\mathrm{H_2}}$. At higher densities, the value of $\gamma_{\rm net}$ ranges from near $5/3$ to near $7/5$, tracking both \hii fraction and temperature. The adiabatic index of the \hii alone, \gammah, has a mean value of 1.39 with a standard deviation of 0.02, but there are excursions to both higher and lower values. Excursions to higher \gammah represent cells that have cooled to near the CMB floor of 85 K, cold enough for the rotational degrees of freedom to freeze out, while those to low \gammah are preferentially cells at temperatures of a few thousand K, where the vibrational degrees of freedom become excited and \gammah reaches an absolute minimum $ = 9/7 \approx 1.28$. 

\subsection{Mass Distribution of Sinks}
\label{s:stat_props}
\begin{figure}
\includegraphics[width=\columnwidth]{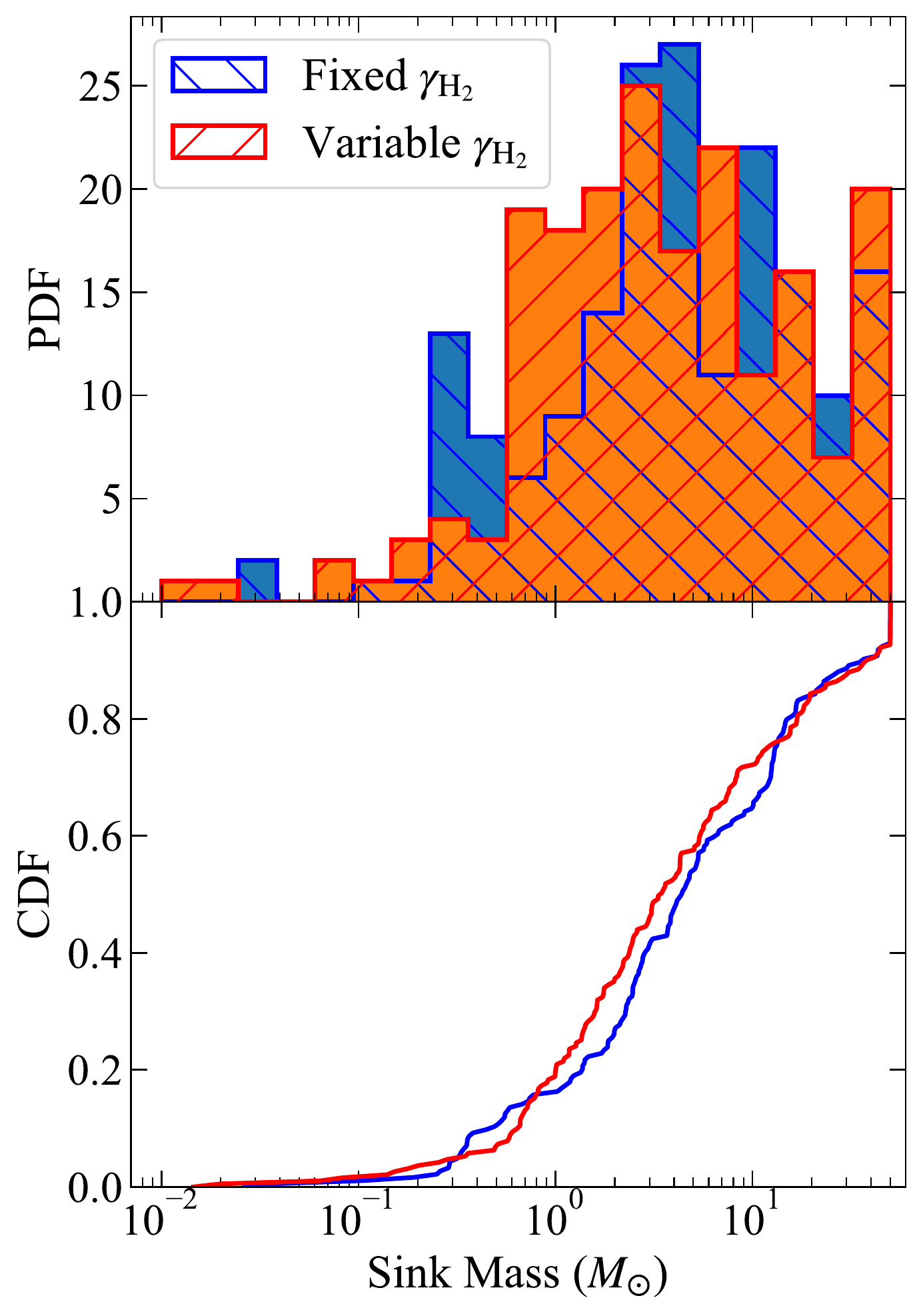}
\caption{Probability distribution function (PDF; \textit{top panel}) and cumulative distribution function (CDF; \textit{bottom panel}) of sink particle masses formed in all 80 simulations. The fixed \gammah case (blue) forms a total of 186 sink particles up to the point where 5 per cent of the initial cloud mass has been deposited in sink particles in each run ($\mathrm{SFE} = 5\%$). The variable \gammah case (red) creates 192 sink particles. Comparing the two distributions with a KS test yields a $p$ value of 0.28, implying that we cannot rule out the null hypothesis that the two sets of sink particle masses were drawn from the same parent distribution. The peak at $50\,\mathrm{M_{\odot}}$ in the PDF and the corresponding jump in the CDF in both sets of runs is due to runs where no fragmentation occurs, and our condition of stopping at $\mathrm{SFE}=5\%$ therefore results in a single sink particle of mass $50\,\mathrm{M_{\odot}}$.}
\label{fig:pdfcdf}
\end{figure}

We next examine the distribution of sink particle masses in the two sets of simulations. The fixed and variable \gammah cases form 186 and 192 sink particles in total, respectively. \autoref{fig:pdfcdf} shows the probability distribution function (PDF) and cumulative distribution function (CDF) of the sink masses at the instant when 5 per cent of the total initial cloud mass has been accreted (\textit{i.e.}, $\mathrm{SFE} = 5\%$), summed over all 80 simulations. We remind the reader that these are not the final masses, since we have not run to 100\% accretion, and do not include the feedback that would be required to do so. However, comparison of early fragmentation in the two simulation sets is nonetheless revealing of whether changes in the H$_2$ adiabatic index matter. We find that the mass distribution peaks around $1\,\mathrm{M_{\odot}}$ in both the fixed and variable $\gamma_{\mathrm{H_2}}$ cases, and rapidly declines for subsolar masses. Both the fixed and the variable \gammah cases have sink particles masses between $0.02-50\,\mathrm{M_{\odot}}$, with a mean of $10.5\,\mathrm{M_{\odot}}$. The two subsets further show quantitatively similar accretion rates of the different sink particles that form in the system. The apparent bi-modality in the distribution caused by the peak at $50\,\mathrm{M_{\odot}}$ is due to the fact that one-third of all the simulations only form a single massive star (no signs of fragmentation until $\mathrm{SFE}=5\%$). In such runs, the single sink particle accretes $50\,\mathrm{M_{\odot}}$. 

To search for differences between the mass distributions for fixed and variable \gammah, we apply the Kolmogorov-Smirnov test (KS-test), which yields a $p$ value of 0.28, implying that we cannot rule out the null hypothesis that the mass distribution is unaffected by our differing treatments of \gammah. Hence, even though the physical properties of the two cases are different (as discussed in \autoref{s:phys_props}), the mass distribution of the sink particles is statistically the same. Of course we cannot rule out the possibility that a difference might become apparent if we performed a larger number of runs, or included feedback enabling the runs to continue further. However, at the level of data available (378 distinct sink particles, measured at $\mathrm{SFE} = 5\%$), changing our treatment of \gammah has no detectable effect.

\subsection{Multiplicity Fraction}
\label{s:mult_frac}

Given that our simulations frequently yield multiple stars (see \autoref{fig:snaps_numdens}), we next examine the multiplicity properties of the stars. A simulation that produces a realistic IMF of the first stars should also be able to explain or predict the fraction of Population III binaries or higher-order bound systems \citep{2009MNRAS.393..663W,2010MNRAS.403...45S}, which is a crucial input to estimates of the rate of black hole or neutron star mergers, and similar high-energy phenomena. 

\begin{figure}
\includegraphics[width=\columnwidth]{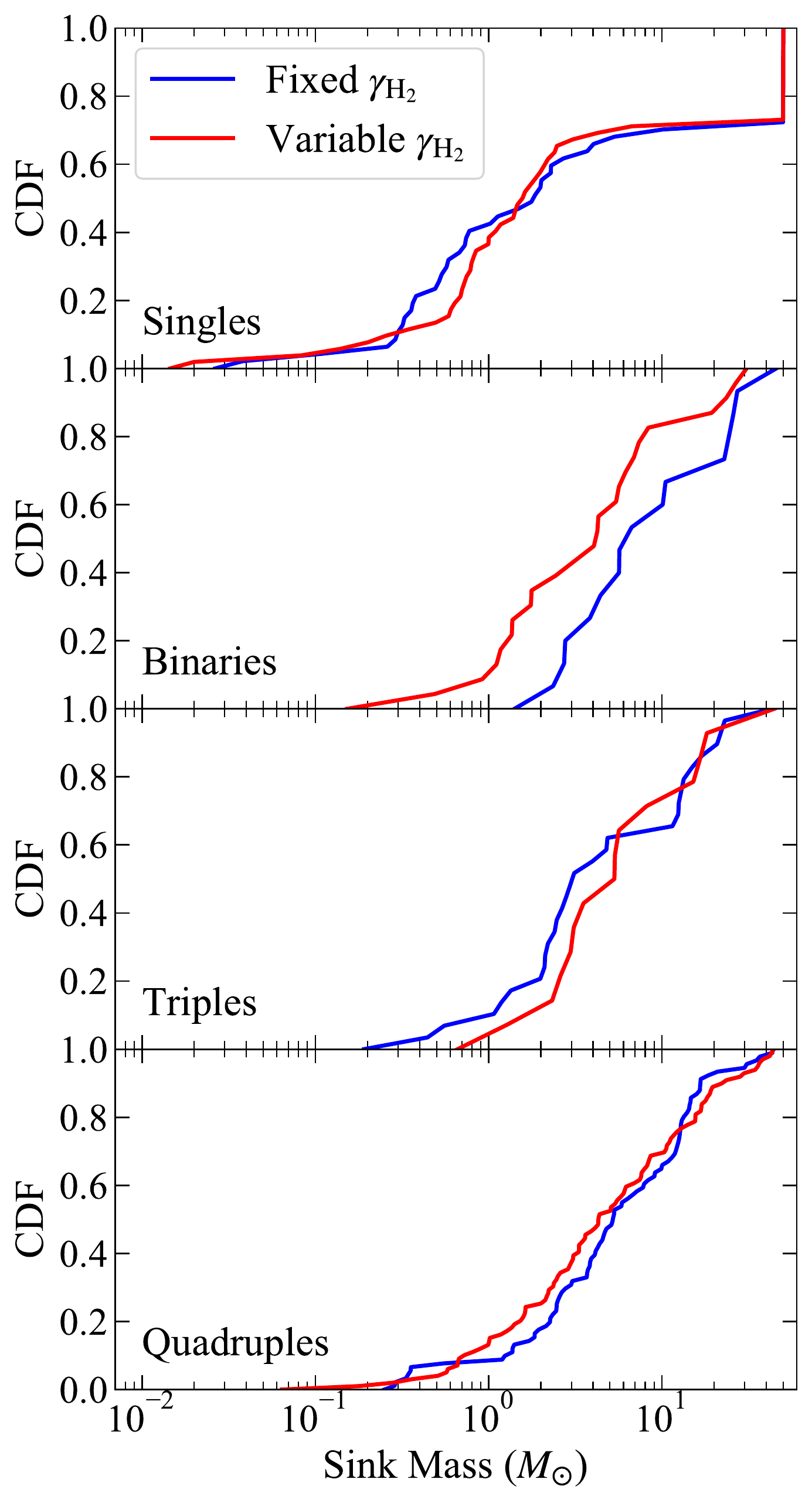}
\caption{Cumulative mass distribution for stars classified as single, binary, triple, and quadruple (top to bottom) in the two cases of fixed and variable \gammah, at a time when 5 per cent of the initial cloud mass has been accreted by sink particles ($\mathrm{SFE} = 5\%$). Stars are classified by multiplicity as described in the main text. The sudden vertical jump at $50\,\mathrm{M_{\odot}}$ in the case of single stars represents the runs that show no fragmentation until $\mathrm{SFE} = 5\%$. Comparisons of the plotted mass distributions via KS tests yields $p$ values consistent with the null hypothesis that both runs are drawn from the same parent distribution.
}

\label{fig:cdf_allles}
\end{figure}
    
\begin{figure}
\includegraphics[width=1.0\columnwidth]{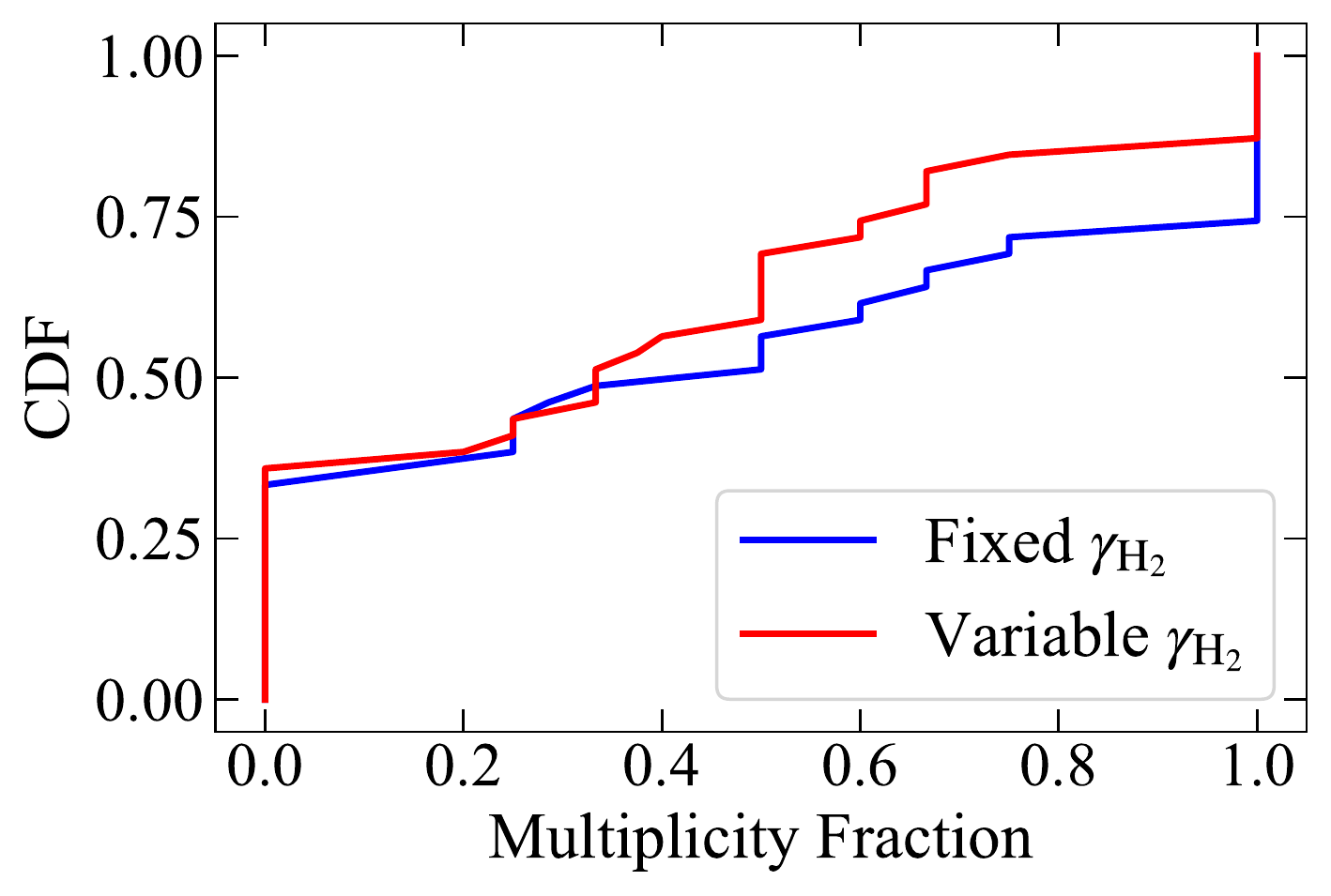}
\caption{CDF of the multiplicity fraction ($mf$) sampled from 80 simulations for the fixed and variable \gammah cases, calculated using \autoref{eq:multfrac}. The high fraction of $mf$ at 0 multiplicity represents one-third of the total runs where no fragmentation is observed. Similarly, runs where all the stars are bound (\textit{i.e.,} no singles) contribute to the jump seen at $mf=1$. The KS-test p-value for the two distributions of $mf$ corresponding to the fixed and variable \gammah cases is 0.72.}
\label{fig:cdfmultfrac}
\end{figure}

\begin{figure}
\includegraphics[width=1.0\columnwidth]{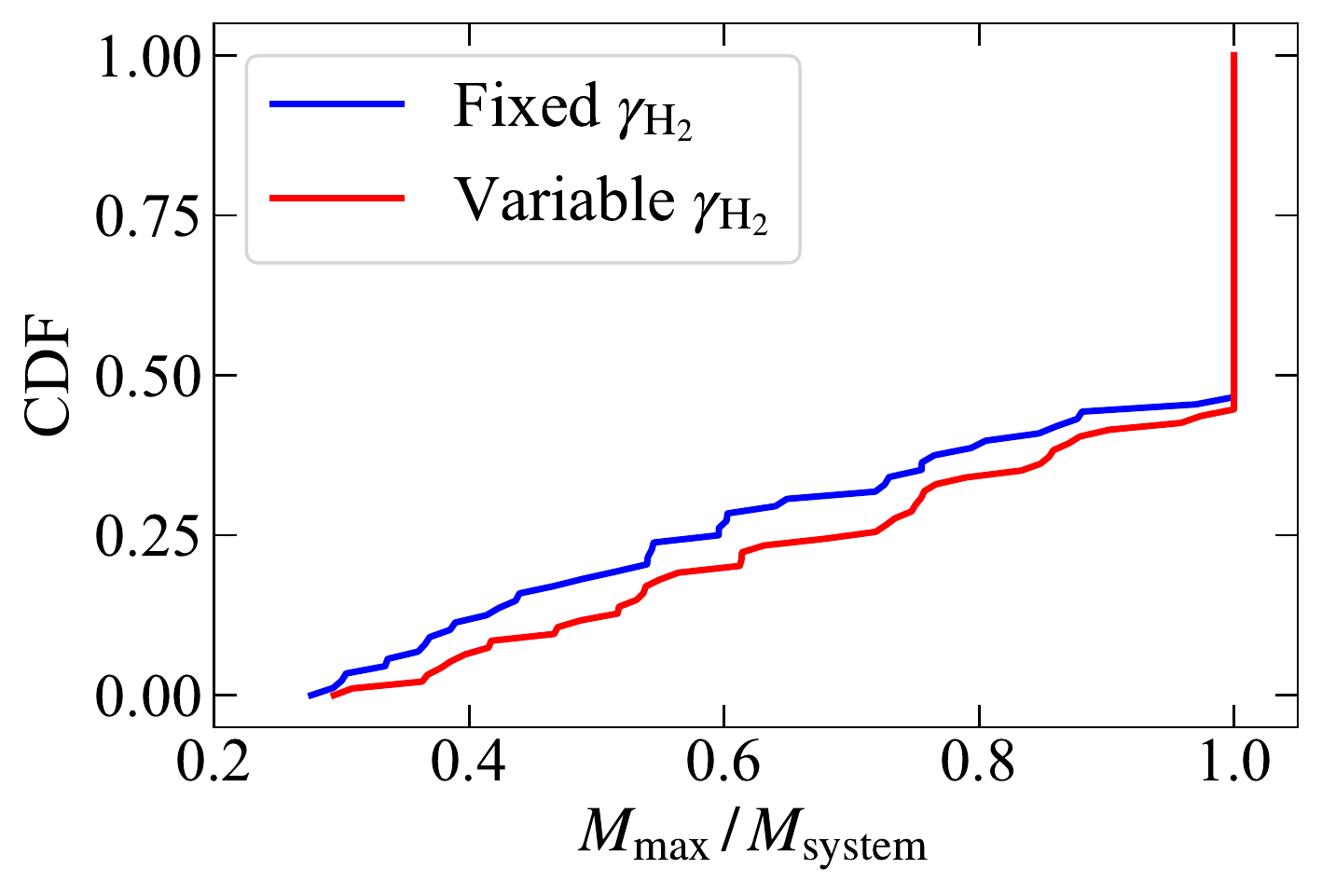}
\caption{CDF of the ratio of the primary (most massive) star ($M_{\mathrm{max}}$) to the sum of masses of stars in a bound system ($M_{\mathrm{system}}$) that can be a binary, triple or quadruple. The substantial fraction of non-fragmented runs lead to this ratio being 1 with a high frequency.}
 \label{fig:cdfmarkratio}
\end{figure}

We define multiplicity in our simulations following the algorithm of \cite{2009MNRAS.392.1363B}. In this algorithm, we recursively find the most bound pair of sinks (\textit{i.e.,} sinks for which the sum of gravitational potential energy and kinetic energy around their common center of mass is most negative) and replace them with a single sink at their center of mass, with mass equal to the sum of masses of the constituent sinks. The velocity of the replaced sink is then the velocity of the center of mass of the constituent pair. Every iteration likewise finds the most bound pair in the system and replaces it with a single sink. This can lead to aggregation of sinks to higher orders, for example, when a sink that replaced two sinks in an earlier iteration now forms the most bound pair with a third sink. The exception to this rule is if aggregating the most bound pair would lead to a quintuple or higher system, which would almost certainly disintegrate dynamically were the simulation to be run long enough; if aggregating the most-bound pair would lead to such an outcome, we skip it and proceed to the next-most-bound pair that can be combined to produce an aggregate of $<5$ individual stars. The algorithm terminates if during any iteration there are no more bound pairs that can be aggregated to yield a combined system with multiplicity $<5$. 

The final result of this algorithm is that all sinks in each simulation are placed in multiplicity groups: isolated sinks that are not bound to any other sinks ($S$), bound pairs ($B$), triples ($T$) or quadruples ($Q$). Then, the multiplicity fraction is given as (for example, \citealt{2010MNRAS.404.1835K,2012MNRAS.419.3115B,2012ApJ...754...71K})
\begin{equation}
\label{eq:multfrac}
\mathrm{mf} = \frac{B + T + Q}{S + B + T + Q}\,.
\end{equation}
This empirical definition has proven robust for use on observations because it does not change if the number of components in a bound system is updated \citep{2005A&A...437..113H}, for example, if a system initially classified as a binary is later discovered to contain a third member. 

\autoref{fig:cdf_allles} shows the CDF of mass for singles, binaries, triples and quadruples in our simulations; note that the CDF we plot is the distribution of masses for \textit{all} stars identified at a given multiplicity, not just for the primary in each system. The (fixed, variable) \gammah cases form (48,$\,$53) single stars, (16,$\,$24) binaries, (30,$\,$15) triples and (92,$\,$100) quadruples, respectively. The mean value of $\mathrm{mf}$ we find for the fixed and variable \gammah cases is 0.45 and 0.37, respectively. Although the differences in the number of binaries and triples for the two cases is 33 and 100 per cent respectively, we cannot classify them as significant because of the low number of such systems that form in our simulations. We compare the fixed and variable \gammah CDFs in each multiplicity bin using a KS test, obtaining $p$ values of 0.43, 0.17, 0.58 and 0.75, for singles, binaries, triples, and quadruples, respectively. As with the mass distribution for all stars, we detect no statistically-significant difference between the fixed and variable \gammah runs for the singles, binaries, triples and quadruples. We speculate that part of the reason that including variable \gammah has little effect is that a cancellation occurs: as shown in \autoref{fig:gammaplot}, depending on the density and temperature regime, values of \gammah both lower and higher than the classical value of 1.4 can occur. The former favours fragmentation (since a gas with lower $\gamma$ is more compressible), while the latter opposes it, but since there are deviations to both lower and higher \gammah the effects roughly cancel out.

We can also compare the multiplicity fractions directly. \autoref{fig:cdfmultfrac} shows the CDF of the multiplicity fraction for the two cases. A KS test comparison of the two distributions yields a $p$ value of 0.72, indicating that the differences in multiplicity fraction are, like the differences in mass, not statistically significant. \autoref{fig:cdfmarkratio} shows the fraction of the mass in multiple systems that is in the primary (most massive star). Values close to unity indicate systems consisting of a dominant primary with small companions, and usually correspond to runs where fragmentation occurs only shortly before we reach $\mbox{SFE}=5\%$, or to cases where fragments form earlier but are not able to accrete much mass. On the other hand, ratios farther from unity indicate more equal multiples, which generally result from near-simultaneous fragmentation at locations very close to each other, although there can be other possibilities. The $p$ value returned by a KS comparison of the variable and fixed \gammah distributions is 0.81, again revealing no statistically-significant differences.

\section{Conclusions}
\label{s:conclusions}

In this work, we study the effects of the adiabatic index of molecular hydrogen (\gammah) on the fragmentation and mass distribution of the first protostars. \hii is the primary component of the star-forming gas and the dominant cooling agent in zero metallicity primordial clouds where there is no dust. Thus, the thermodynamics are primarily controlled by \hii and as such it is necessary to check whether the common approximation of treating \hii as a classical diatomic gas with fixed adiabatic index \gammah = 7/5 is valid. The reason for concern is that, at the temperatures of a few hundred K found in primordial clouds and the accretion discs of the first stars, rotational and vibrational levels of \hii are only starting to become excited, and thus \hii behaves non-classically. Instead, its adiabatic index is a complex function of both temperature and the ratio of ortho- to para-\hii. 

We perform two sets of 3D simulations at high resolution ($7.6\,\mathrm{AU}$) using the AMR hydrodynamic code FLASH. In the first set we fix $\gamma_{\mathrm{H_2}} = 7/5$, and in the second we compute \gammah as a function of temperature and ortho- to para- ratio using a full quantum mechanical treatment. We follow all simulations up to the time when 5 per cent of the initial cloud mass is accreted by sink particles, yielding a total of 378 sink particles from 80 simulations with different initial random turbulent velocity fields, with a mean sink particle mass of $10.5\,\mathrm{M_{\odot}}$. We show that primordial systems can undergo high fragmentation at distances as close as $5\,\mathrm{AU}$ from the primary star, thus resulting in stars being bound to others soon after their formation; the mean multiplicity fraction is 0.4. However, around one-third of our simulations show no fragmentation even when the first star has accreted $50\,\mathrm{M_{\odot}}$. Hence, unless feedback effectively halts accretion on to the isolated massive stars, it seems likely that a great diversity of Population III stars existed, from single massive stars evolving in isolation to clustered formation of significantly less massive stars. 

Comparing runs using identical initial conditions run with fixed and variable \gammah reveals clear differences in physical properties such as density and temperature distributions, especially in regions where the net adiabatic index of all species ($\gamma_{\mathrm{net}}$) is dominated by \gammah due to the high mass fraction of \hii. We compare and analyze the mass distribution and multiplicity of the fixed and variable \gammah simulations; we find no statistically significant differences between the two. We also do not see any evidence of preferential formation of clustered systems in either of the two cases. Within the domains covered by this analysis, we therefore conclude that the standard approximation of molecular hydrogen as a classical diatomic gas during the formation of the first stars is valid, at least to first order during the first few thousand years after collapse of the formation of the first star. This may change with feedback, as feedback will alter the temperature distribution and hence the value of \gammah. Thus it is possible that a full quantum mechanical treatment of the \hii molecule will prove more important at later stages of the primordial star formation process. Nonetheless, we recommend following the accurate treatment of the \hii adiabatic index in future studies of formation of the first stars because it is not computationally more expensive as compared to the classical treatment.

\section*{Acknowledgements}
We thank the referee Naoki Yoshida for a positive and encouraging report that helped improve the presentation of our results. We thank Chris Power for computing resources to run the convergence simulations. PS is supported by an Australian Government Research Training Program (RTP) Scholarship. MRK and CF acknowledge funding provided by the Australian Research Council (ARC) through Discovery Projects DP190101258 (MRK) and DP170100603 (CF) and Future Fellowships FT180100375 (MRK) and FT180100495 (CF), and the Australia-Germany Joint Research Cooperation Scheme (UA-DAAD; both MRK and CF). 
    
The simulations and data analyses presented in this work used high performance computing resources provided by the Australian National Computational Infrastructure (NCI) through projects \texttt{ek9} (CF) and \texttt{jh2} (MRK) in the framework of the National Computational Merit Allocation Scheme and the Australian National University (ANU) Allocation Scheme, and as part of contribution by NCI to the ARC Centre of Excellence for All Sky Astrophysics in 3 Dimensions (ASTRO 3D, CE170100013). The simulation software FLASH was in part developed by the DOE-supported Flash Centre for Computational Science at the University of Chicago. Analysis was performed using \texttt{yt} \citep{2011ApJS..192....9T}.
    
    

    
\bibliographystyle{mnras}
\bibliography{references} 

\begin{thebibliography}{}
\makeatletter
\relax
\def\mn@urlcharsother{\let\do\@makeother \do\$\do\&\do\#\do\^\do\_\do\%\do\~}
\def\mn@doi{\begingroup\mn@urlcharsother \@ifnextchar [ {\mn@doi@}
  {\mn@doi@[]}}
\def\mn@doi@[#1]#2{\def\@tempa{#1}\ifx\@tempa\@empty \href
  {http://dx.doi.org/#2} {doi:#2}\else \href {http://dx.doi.org/#2} {#1}\fi
  \endgroup}
\def\mn@eprint#1#2{\mn@eprint@#1:#2::\@nil}
\def\mn@eprint@arXiv#1{\href {http://arxiv.org/abs/#1} {{\tt arXiv:#1}}}
\def\mn@eprint@dblp#1{\href {http://dblp.uni-trier.de/rec/bibtex/#1.xml}
  {dblp:#1}}
\def\mn@eprint@#1:#2:#3:#4\@nil{\def\@tempa {#1}\def\@tempb {#2}\def\@tempc
  {#3}\ifx \@tempc \@empty \let \@tempc \@tempb \let \@tempb \@tempa \fi \ifx
  \@tempb \@empty \def\@tempb {arXiv}\fi \@ifundefined
  {mn@eprint@\@tempb}{\@tempb:\@tempc}{\expandafter \expandafter \csname
  mn@eprint@\@tempb\endcsname \expandafter{\@tempc}}}

\bibitem[\protect\citeauthoryear{{Abel}, {Bryan}  \& {Norman}}{{Abel}
  et~al.}{2000}]{2000ApJ...540...39A}
{Abel} T.,  {Bryan} G.~L.,   {Norman} M.~L.,  2000, \mn@doi [\apj]
  {10.1086/309295}, \href
  {https://ui.adsabs.harvard.edu/abs/2000ApJ...540...39A} {540, 39}

\bibitem[\protect\citeauthoryear{{Abel}, {Bryan}  \& {Norman}}{{Abel}
  et~al.}{2002}]{2002Sci...295...93A}
{Abel} T.,  {Bryan} G.~L.,   {Norman} M.~L.,  2002, \mn@doi [Science]
  {10.1126/science.295.5552.93}, \href
  {http://adsabs.harvard.edu/abs/2002Sci...295...93A} {295, 93}

\bibitem[\protect\citeauthoryear{{Ahn} \& {Shapiro}}{{Ahn} \&
  {Shapiro}}{2007}]{2007MNRAS.375..881A}
{Ahn} K.,  {Shapiro} P.~R.,  2007, \mn@doi [\mnras]
  {10.1111/j.1365-2966.2006.11332.x}, \href
  {https://ui.adsabs.harvard.edu/abs/2007MNRAS.375..881A} {375, 881}

\bibitem[\protect\citeauthoryear{{Bate}}{{Bate}}{1998}]{1998ApJ...508L..95B}
{Bate} M.~R.,  1998, \mn@doi [\apjl] {10.1086/311719}, \href
  {http://adsabs.harvard.edu/abs/1998ApJ...508L..95B} {508, L95}

\bibitem[\protect\citeauthoryear{{Bate}}{{Bate}}{2009}]{2009MNRAS.392.1363B}
{Bate} M.~R.,  2009, \mn@doi [\mnras] {10.1111/j.1365-2966.2008.14165.x}, \href
  {https://ui.adsabs.harvard.edu/abs/2009MNRAS.392.1363B} {392, 1363}

\bibitem[\protect\citeauthoryear{{Bate}}{{Bate}}{2012}]{2012MNRAS.419.3115B}
{Bate} M.~R.,  2012, \mn@doi [\mnras] {10.1111/j.1365-2966.2011.19955.x}, \href
  {https://ui.adsabs.harvard.edu/abs/2012MNRAS.419.3115B} {419, 3115}

\bibitem[\protect\citeauthoryear{{Bate}, {Bonnell}  \& {Price}}{{Bate}
  et~al.}{1995}]{1995MNRAS.277..362B}
{Bate} M.~R.,  {Bonnell} I.~A.,   {Price} N.~M.,  1995, \mn@doi [\mnras]
  {10.1093/mnras/277.2.362}, \href
  {http://adsabs.harvard.edu/abs/1995MNRAS.277..362B} {277, 362}

\bibitem[\protect\citeauthoryear{{Berger} \& {Colella}}{{Berger} \&
  {Colella}}{1989}]{1989JCoPh..82...64B}
{Berger} M.~J.,  {Colella} P.,  1989, \mn@doi [Journal of Computational
  Physics] {10.1016/0021-9991(89)90035-1}, \href
  {http://adsabs.harvard.edu/abs/1989JCoPh..82...64B} {82, 64}

\bibitem[\protect\citeauthoryear{{Bitsch}, {Boley}  \& {Kley}}{{Bitsch}
  et~al.}{2013}]{2013A&A...550A..52B}
{Bitsch} B.,  {Boley} A.,   {Kley} W.,  2013, \mn@doi [\aap]
  {10.1051/0004-6361/201118490}, \href
  {https://ui.adsabs.harvard.edu/abs/2013A%26A...550A..52B} {550, A52}

\bibitem[\protect\citeauthoryear{{Black} \& {Bodenheimer}}{{Black} \&
  {Bodenheimer}}{1975}]{1975ApJ...199..619B}
{Black} D.~C.,  {Bodenheimer} P.,  1975, \mn@doi [\apj] {10.1086/153729}, \href
  {http://adsabs.harvard.edu/abs/1975ApJ...199..619B} {199, 619}

\bibitem[\protect\citeauthoryear{{Bleuler} \& {Teyssier}}{{Bleuler} \&
  {Teyssier}}{2014}]{2014MNRAS.445.4015B}
{Bleuler} A.,  {Teyssier} R.,  2014, \mn@doi [\mnras] {10.1093/mnras/stu2005},
  \href {http://adsabs.harvard.edu/abs/2014MNRAS.445.4015B} {445, 4015}

\bibitem[\protect\citeauthoryear{{Boley}, {Hartquist}, {Durisen}  \&
  {Michael}}{{Boley} et~al.}{2007}]{2007ApJ...656L..89B}
{Boley} A.~C.,  {Hartquist} T.~W.,  {Durisen} R.~H.,   {Michael} S.,  2007,
  \mn@doi [\apjl] {10.1086/512235}, \href
  {http://adsabs.harvard.edu/abs/2007ApJ...656L..89B} {656, L89}

\bibitem[\protect\citeauthoryear{{Bond}}{{Bond}}{1981}]{1981ApJ...248..606B}
{Bond} H.~E.,  1981, \mn@doi [\apj] {10.1086/159186}, \href
  {http://adsabs.harvard.edu/abs/1981ApJ...248..606B} {248, 606}

\bibitem[\protect\citeauthoryear{Bouchut, Klingenberg  \& Waagan}{Bouchut
  et~al.}{2007}]{Bouchut2007}
Bouchut F.,  Klingenberg C.,   Waagan K.,  2007, \mn@doi [Numerische
  Mathematik] {10.1007/s00211-007-0108-8}, 108, 7

\bibitem[\protect\citeauthoryear{Bouchut, Klingenberg  \& Waagan}{Bouchut
  et~al.}{2010}]{Bouchut2010}
Bouchut F.,  Klingenberg C.,   Waagan K.,  2010, \mn@doi [Numerische
  Mathematik] {10.1007/s00211-010-0289-4}, 115, 647

\bibitem[\protect\citeauthoryear{{Bovino}, {Schleicher}  \& {Schober}}{{Bovino}
  et~al.}{2013a}]{2013NJPh...15a3055B}
{Bovino} S.,  {Schleicher} D.~R.~G.,   {Schober} J.,  2013a, \mn@doi [New
  Journal of Physics] {10.1088/1367-2630/15/1/013055}, \href
  {https://ui.adsabs.harvard.edu/abs/2013NJPh...15a3055B} {15, 013055}

\bibitem[\protect\citeauthoryear{{Bovino}, {Grassi}, {Latif}  \&
  {Schleicher}}{{Bovino} et~al.}{2013b}]{2013MNRAS.434L..36B}
{Bovino} S.,  {Grassi} T.,  {Latif} M.~A.,   {Schleicher} D.~R.~G.,  2013b,
  \mn@doi [\mnras] {10.1093/mnrasl/slt072}, \href
  {http://adsabs.harvard.edu/abs/2013MNRAS.434L..36B} {434, L36}

\bibitem[\protect\citeauthoryear{{Brandenburg}, {Enqvist}  \&
  {Olesen}}{{Brandenburg} et~al.}{1996}]{1996PhRvD..54.1291B}
{Brandenburg} A.,  {Enqvist} K.,   {Olesen} P.,  1996, \mn@doi [\prd]
  {10.1103/PhysRevD.54.1291}, \href
  {https://ui.adsabs.harvard.edu/abs/1996PhRvD..54.1291B} {54, 1291}

\bibitem[\protect\citeauthoryear{{Brandenburg}, {Sokoloff}  \&
  {Subramanian}}{{Brandenburg} et~al.}{2012}]{2012SSRv..169..123B}
{Brandenburg} A.,  {Sokoloff} D.,   {Subramanian} K.,  2012, \mn@doi [\ssr]
  {10.1007/s11214-012-9909-x}, \href
  {http://adsabs.harvard.edu/abs/2012SSRv..169..123B} {169, 123}

\bibitem[\protect\citeauthoryear{{Bromm}}{{Bromm}}{2013}]{2013RPPh...76k2901B}
{Bromm} V.,  2013, \mn@doi [Reports on Progress in Physics]
  {10.1088/0034-4885/76/11/112901}, \href
  {https://ui.adsabs.harvard.edu/abs/2013RPPh...76k2901B} {76, 112901}

\bibitem[\protect\citeauthoryear{{Bromm} \& {Larson}}{{Bromm} \&
  {Larson}}{2004}]{2004ARA&A..42...79B}
{Bromm} V.,  {Larson} R.~B.,  2004, \mn@doi [\araa]
  {10.1146/annurev.astro.42.053102.134034}, \href
  {http://adsabs.harvard.edu/abs/2004ARA%26A..42...79B} {42, 79}

\bibitem[\protect\citeauthoryear{{Bromm}, {Coppi}  \& {Larson}}{{Bromm}
  et~al.}{2002}]{2002ApJ...564...23B}
{Bromm} V.,  {Coppi} P.~S.,   {Larson} R.~B.,  2002, \mn@doi [\apj]
  {10.1086/323947}, \href {http://adsabs.harvard.edu/abs/2002ApJ...564...23B}
  {564, 23}

\bibitem[\protect\citeauthoryear{{Burkert} \& {Bodenheimer}}{{Burkert} \&
  {Bodenheimer}}{2000}]{2000ApJ...543..822B}
{Burkert} A.,  {Bodenheimer} P.,  2000, \mn@doi [\apj] {10.1086/317122}, \href
  {https://ui.adsabs.harvard.edu/abs/2000ApJ...543..822B} {543, 822}

\bibitem[\protect\citeauthoryear{{Ciardi} \& {Ferrara}}{{Ciardi} \&
  {Ferrara}}{2005}]{2005SSRv..116..625C}
{Ciardi} B.,  {Ferrara} A.,  2005, \mn@doi [\ssr] {10.1007/s11214-005-3592-0},
  \href {http://adsabs.harvard.edu/abs/2005SSRv..116..625C} {116, 625}

\bibitem[\protect\citeauthoryear{{Clark}, {Glover}, {Smith}, {Greif}, {Klessen}
   \& {Bromm}}{{Clark} et~al.}{2011a}]{2011Sci...331.1040C}
{Clark} P.~C.,  {Glover} S.~C.~O.,  {Smith} R.~J.,  {Greif} T.~H.,  {Klessen}
  R.~S.,   {Bromm} V.,  2011a, \mn@doi [Science] {10.1126/science.1198027},
  \href {http://adsabs.harvard.edu/abs/2011Sci...331.1040C} {331, 1040}

\bibitem[\protect\citeauthoryear{{Clark}, {Glover}, {Klessen}  \&
  {Bromm}}{{Clark} et~al.}{2011b}]{2011ApJ...727..110C}
{Clark} P.~C.,  {Glover} S.~C.~O.,  {Klessen} R.~S.,   {Bromm} V.,  2011b,
  \mn@doi [\apj] {10.1088/0004-637X/727/2/110}, \href
  {https://ui.adsabs.harvard.edu/abs/2011ApJ...727..110C} {727, 110}

\bibitem[\protect\citeauthoryear{{Commer{\c{c}}on}, {Hennebelle}, {Audit},
  {Chabrier}  \& {Teyssier}}{{Commer{\c{c}}on}
  et~al.}{2008}]{2008A&A...482..371C}
{Commer{\c{c}}on} B.,  {Hennebelle} P.,  {Audit} E.,  {Chabrier} G.,
  {Teyssier} R.,  2008, \mn@doi [\aap] {10.1051/0004-6361:20078591}, \href
  {https://ui.adsabs.harvard.edu/abs/2008A&A...482..371C} {482, 371}

\bibitem[\protect\citeauthoryear{{Cornuault}, {Lehnert}, {Boulanger}  \&
  {Guillard}}{{Cornuault} et~al.}{2018}]{2018A&A...610A..75C}
{Cornuault} N.,  {Lehnert} M.~D.,  {Boulanger} F.,   {Guillard} P.,  2018,
  \mn@doi [\aap] {10.1051/0004-6361/201629229}, \href
  {https://ui.adsabs.harvard.edu/abs/2018A&A...610A..75C} {610, A75}

\bibitem[\protect\citeauthoryear{{De Souza}, {Yoshida}  \& {Ioka}}{{De Souza}
  et~al.}{2011}]{2011A&A...533A..32D}
{De Souza} R.~S.,  {Yoshida} N.,   {Ioka} K.,  2011, \mn@doi [\aap]
  {10.1051/0004-6361/201117242}, \href
  {https://ui.adsabs.harvard.edu/abs/2011A&A...533A..32D} {533, A32}

\bibitem[\protect\citeauthoryear{{Draine}, {Roberge}  \& {Dalgarno}}{{Draine}
  et~al.}{1983}]{1983ApJ...264..485D}
{Draine} B.~T.,  {Roberge} W.~G.,   {Dalgarno} A.,  1983, \mn@doi [\apj]
  {10.1086/160617}, \href
  {https://ui.adsabs.harvard.edu/abs/1983ApJ...264..485D} {264, 485}

\bibitem[\protect\citeauthoryear{{Dubey} et~al.,}{{Dubey}
  et~al.}{2008}]{2008ASPC..385..145D}
{Dubey} A.,  et~al., 2008, in {Pogorelov} N.~V.,  {Audit} E.,   {Zank} G.~P.,
  eds,  Astronomical Society of the Pacific Conference Series Vol. 385,
  Numerical Modeling of Space Plasma Flows. p.~145

\bibitem[\protect\citeauthoryear{{Epstein}, {Lattimer}  \& {Schramm}}{{Epstein}
  et~al.}{1976}]{1976Natur.263..198E}
{Epstein} R.~I.,  {Lattimer} J.~M.,   {Schramm} D.~N.,  1976, \mn@doi [\nat]
  {10.1038/263198a0}, \href
  {https://ui.adsabs.harvard.edu/abs/1976Natur.263..198E} {263, 198}

\bibitem[\protect\citeauthoryear{{Federrath}}{{Federrath}}{2013}]{2013MNRAS.436.1245F}
{Federrath} C.,  2013, \mn@doi [\mnras] {10.1093/mnras/stt1644}, \href
  {https://ui.adsabs.harvard.edu/abs/2013MNRAS.436.1245F} {436, 1245}

\bibitem[\protect\citeauthoryear{{Federrath}, {Banerjee}, {Clark}  \&
  {Klessen}}{{Federrath} et~al.}{2010}]{2010ApJ...713..269F}
{Federrath} C.,  {Banerjee} R.,  {Clark} P.~C.,   {Klessen} R.~S.,  2010,
  \mn@doi [\apj] {10.1088/0004-637X/713/1/269}, \href
  {https://ui.adsabs.harvard.edu/\#abs/2010ApJ...713..269F} {713, 269}

\bibitem[\protect\citeauthoryear{{Federrath}, {Banerjee}, {Seifried}, {Clark}
  \& {Klessen}}{{Federrath} et~al.}{2011a}]{2011IAUS..270..425F}
{Federrath} C.,  {Banerjee} R.,  {Seifried} D.,  {Clark} P.~C.,   {Klessen}
  R.~S.,  2011a, in {Alves} J.,  {Elmegreen} B.~G.,  {Girart} J.~M.,
  {Trimble} V.,  eds,  IAU Symposium Vol. 270, Computational Star Formation. pp
  425--428 (\mn@eprint {arXiv} {1007.2504}), \mn@doi{10.1017/S1743921311000755}

\bibitem[\protect\citeauthoryear{{Federrath}, {Sur}, {Schleicher}, {Banerjee}
  \& {Klessen}}{{Federrath} et~al.}{2011b}]{2011ApJ...731...62F}
{Federrath} C.,  {Sur} S.,  {Schleicher} D.~R.~G.,  {Banerjee} R.,   {Klessen}
  R.~S.,  2011b, \mn@doi [\apj] {10.1088/0004-637X/731/1/62}, \href
  {http://adsabs.harvard.edu/abs/2011ApJ...731...62F} {731, 62}

\bibitem[\protect\citeauthoryear{{Federrath}, {Schober}, {Bovino}  \&
  {Schleicher}}{{Federrath} et~al.}{2014}]{2014ApJ...797L..19F}
{Federrath} C.,  {Schober} J.,  {Bovino} S.,   {Schleicher} D.~R.~G.,  2014,
  \mn@doi [\apjl] {10.1088/2041-8205/797/2/L19}, \href
  {http://adsabs.harvard.edu/abs/2014ApJ...797L..19F} {797, L19}

\bibitem[\protect\citeauthoryear{{Fialkov}, {Barkana}  \& {Visbal}}{{Fialkov}
  et~al.}{2014}]{2014Natur.506..197F}
{Fialkov} A.,  {Barkana} R.,   {Visbal} E.,  2014, \mn@doi [\nat]
  {10.1038/nature12999}, \href
  {http://adsabs.harvard.edu/abs/2014Natur.506..197F} {506, 197}

\bibitem[\protect\citeauthoryear{{Fields}}{{Fields}}{2011}]{2011ARNPS..61...47F}
{Fields} B.~D.,  2011, \mn@doi [Annual Review of Nuclear and Particle Science]
  {10.1146/annurev-nucl-102010-130445}, \href
  {https://ui.adsabs.harvard.edu/abs/2011ARNPS..61...47F} {61, 47}

\bibitem[\protect\citeauthoryear{{Flower} \& {Harris}}{{Flower} \&
  {Harris}}{2007}]{2007MNRAS.377..705F}
{Flower} D.~R.,  {Harris} G.~J.,  2007, \mn@doi [\mnras]
  {10.1111/j.1365-2966.2007.11632.x}, \href
  {https://ui.adsabs.harvard.edu/abs/2007MNRAS.377..705F} {377, 705}

\bibitem[\protect\citeauthoryear{{Flower} \& {Pineau des For{\^e}ts}}{{Flower}
  \& {Pineau des For{\^e}ts}}{2000}]{2000MNRAS.316..901F}
{Flower} D.~R.,  {Pineau des For{\^e}ts} G.,  2000, \mn@doi [\mnras]
  {10.1046/j.1365-8711.2000.03611.x}, \href
  {https://ui.adsabs.harvard.edu/abs/2000MNRAS.316..901F} {316, 901}

\bibitem[\protect\citeauthoryear{{Fryxell} et~al.,}{{Fryxell}
  et~al.}{2000}]{2000ApJS..131..273F}
{Fryxell} B.,  et~al., 2000, \mn@doi [\apjs] {10.1086/317361}, \href
  {http://adsabs.harvard.edu/abs/2000ApJS..131..273F} {131, 273}

\bibitem[\protect\citeauthoryear{{Galli} \& {Palla}}{{Galli} \&
  {Palla}}{1998}]{1998A&A...335..403G}
{Galli} D.,  {Palla} F.,  1998, \aap, \href
  {http://adsabs.harvard.edu/abs/1998A%26A...335..403G} {335, 403}

\bibitem[\protect\citeauthoryear{{Galli} \& {Palla}}{{Galli} \&
  {Palla}}{2002}]{2002P&SS...50.1197G}
{Galli} D.,  {Palla} F.,  2002, \mn@doi [\planss]
  {10.1016/S0032-0633(02)00083-1}, \href
  {http://adsabs.harvard.edu/abs/2002P%26SS...50.1197G} {50, 1197}

\bibitem[\protect\citeauthoryear{{Galli} \& {Palla}}{{Galli} \&
  {Palla}}{2013}]{2013ARA&A..51..163G}
{Galli} D.,  {Palla} F.,  2013, \mn@doi [\araa]
  {10.1146/annurev-astro-082812-141029}, \href
  {http://adsabs.harvard.edu/abs/2013ARA%26A..51..163G} {51, 163}

\bibitem[\protect\citeauthoryear{{Ge} \& {Wise}}{{Ge} \&
  {Wise}}{2017}]{2017MNRAS.472.2773G}
{Ge} Q.,  {Wise} J.~H.,  2017, \mn@doi [\mnras] {10.1093/mnras/stx2074}, \href
  {https://ui.adsabs.harvard.edu/abs/2017MNRAS.472.2773G} {472, 2773}

\bibitem[\protect\citeauthoryear{{Girichidis}, {Federrath}, {Banerjee}  \&
  {Klessen}}{{Girichidis} et~al.}{2012}]{2012MNRAS.420..613G}
{Girichidis} P.,  {Federrath} C.,  {Banerjee} R.,   {Klessen} R.~S.,  2012,
  \mn@doi [\mnras] {10.1111/j.1365-2966.2011.20073.x}, \href
  {https://ui.adsabs.harvard.edu/abs/2012MNRAS.420..613G} {420, 613}

\bibitem[\protect\citeauthoryear{{Glover}}{{Glover}}{2005}]{2005SSRv..117..445G}
{Glover} S.,  2005, \mn@doi [\ssr] {10.1007/s11214-005-5821-y}, \href
  {http://adsabs.harvard.edu/abs/2005SSRv..117..445G} {117, 445}

\bibitem[\protect\citeauthoryear{{Glover} \& {Abel}}{{Glover} \&
  {Abel}}{2008}]{2008MNRAS.388.1627G}
{Glover} S.~C.~O.,  {Abel} T.,  2008, \mn@doi [\mnras]
  {10.1111/j.1365-2966.2008.13224.x}, \href
  {http://adsabs.harvard.edu/abs/2008MNRAS.388.1627G} {388, 1627}

\bibitem[\protect\citeauthoryear{{Gong} \& {Ostriker}}{{Gong} \&
  {Ostriker}}{2013}]{2013ApJS..204....8G}
{Gong} H.,  {Ostriker} E.~C.,  2013, \mn@doi [\apjs]
  {10.1088/0067-0049/204/1/8}, \href
  {http://adsabs.harvard.edu/abs/2013ApJS..204....8G} {204, 8}

\bibitem[\protect\citeauthoryear{{Goodman}, {Benson}, {Fuller}  \&
  {Myers}}{{Goodman} et~al.}{1993}]{1993ApJ...406..528G}
{Goodman} A.~A.,  {Benson} P.~J.,  {Fuller} G.~A.,   {Myers} P.~C.,  1993,
  \mn@doi [\apj] {10.1086/172465}, \href
  {https://ui.adsabs.harvard.edu/abs/1993ApJ...406..528G} {406, 528}

\bibitem[\protect\citeauthoryear{{Grassi}, {Bovino}, {Schleicher}  \&
  {Gianturco}}{{Grassi} et~al.}{2013}]{2013MNRAS.431.1659G}
{Grassi} T.,  {Bovino} S.,  {Schleicher} D.,   {Gianturco} F.~A.,  2013,
  \mn@doi [\mnras] {10.1093/mnras/stt284}, \href
  {http://adsabs.harvard.edu/abs/2013MNRAS.431.1659G} {431, 1659}

\bibitem[\protect\citeauthoryear{{Grassi}, {Bovino}, {Schleicher}, {Prieto},
  {Seifried}, {Simoncini}  \& {Gianturco}}{{Grassi}
  et~al.}{2014}]{2014MNRAS.439.2386G}
{Grassi} T.,  {Bovino} S.,  {Schleicher} D.~R.~G.,  {Prieto} J.,  {Seifried}
  D.,  {Simoncini} E.,   {Gianturco} F.~A.,  2014, \mn@doi [\mnras]
  {10.1093/mnras/stu114}, \href
  {http://adsabs.harvard.edu/abs/2014MNRAS.439.2386G} {439, 2386}

\bibitem[\protect\citeauthoryear{{Greif}}{{Greif}}{2014}]{2014MNRAS.444.1566G}
{Greif} T.~H.,  2014, \mn@doi [\mnras] {10.1093/mnras/stu1532}, \href
  {https://ui.adsabs.harvard.edu/abs/2014MNRAS.444.1566G} {444, 1566}

\bibitem[\protect\citeauthoryear{{Greif}, {Johnson}, {Klessen}  \&
  {Bromm}}{{Greif} et~al.}{2008}]{2008MNRAS.387.1021G}
{Greif} T.~H.,  {Johnson} J.~L.,  {Klessen} R.~S.,   {Bromm} V.,  2008, \mn@doi
  [\mnras] {10.1111/j.1365-2966.2008.13326.x}, \href
  {https://ui.adsabs.harvard.edu/abs/2008MNRAS.387.1021G} {387, 1021}

\bibitem[\protect\citeauthoryear{{Griffen}, {Dooley}, {Ji}, {O'Shea},
  {G{\'o}mez}  \& {Frebel}}{{Griffen} et~al.}{2018}]{2018MNRAS.474..443G}
{Griffen} B.~F.,  {Dooley} G.~A.,  {Ji} A.~P.,  {O'Shea} B.~W.,  {G{\'o}mez}
  F.~A.,   {Frebel} A.,  2018, \mn@doi [\mnras] {10.1093/mnras/stx2749}, \href
  {http://adsabs.harvard.edu/abs/2018MNRAS.474..443G} {474, 443}

\bibitem[\protect\citeauthoryear{{Hartwig}, {Clark}, {Glover}, {Klessen}  \&
  {Sasaki}}{{Hartwig} et~al.}{2015}]{2015ApJ...799..114H}
{Hartwig} T.,  {Clark} P.~C.,  {Glover} S. C.~O.,  {Klessen} R.~S.,   {Sasaki}
  M.,  2015, \mn@doi [\apj] {10.1088/0004-637X/799/2/114}, \href
  {https://ui.adsabs.harvard.edu/abs/2015ApJ...799..114H} {799, 114}

\bibitem[\protect\citeauthoryear{{Hartwig}, {Ishigaki}, {Klessen}  \&
  {Yoshida}}{{Hartwig} et~al.}{2019}]{2019MNRAS.482.1204H}
{Hartwig} T.,  {Ishigaki} M.~N.,  {Klessen} R.~S.,   {Yoshida} N.,  2019,
  \mn@doi [\mnras] {10.1093/mnras/sty2783}, \href
  {http://adsabs.harvard.edu/abs/2019MNRAS.482.1204H} {482, 1204}

\bibitem[\protect\citeauthoryear{Hindmarsh}{Hindmarsh}{1980}]{Hindmarsh:1980:LLT:1218052.1218054}
Hindmarsh A.~C.,  1980, \mn@doi [SIGNUM Newsl.] {10.1145/1218052.1218054}, 15,
  10

\bibitem[\protect\citeauthoryear{{Hirano} \& {Yoshida}}{{Hirano} \&
  {Yoshida}}{2013}]{2013ApJ...763...52H}
{Hirano} S.,  {Yoshida} N.,  2013, \mn@doi [\apj] {10.1088/0004-637X/763/1/52},
  \href {http://adsabs.harvard.edu/abs/2013ApJ...763...52H} {763, 52}

\bibitem[\protect\citeauthoryear{{Hirano}, {Hosokawa}, {Yoshida}, {Umeda},
  {Omukai}, {Chiaki}  \& {Yorke}}{{Hirano} et~al.}{2014}]{2014ApJ...781...60H}
{Hirano} S.,  {Hosokawa} T.,  {Yoshida} N.,  {Umeda} H.,  {Omukai} K.,
  {Chiaki} G.,   {Yorke} H.~W.,  2014, \mn@doi [\apj]
  {10.1088/0004-637X/781/2/60}, \href
  {http://adsabs.harvard.edu/abs/2014ApJ...781...60H} {781, 60}

\bibitem[\protect\citeauthoryear{{Hirano}, {Hosokawa}, {Yoshida}, {Omukai}  \&
  {Yorke}}{{Hirano} et~al.}{2015}]{2015MNRAS.448..568H}
{Hirano} S.,  {Hosokawa} T.,  {Yoshida} N.,  {Omukai} K.,   {Yorke} H.~W.,
  2015, \mn@doi [\mnras] {10.1093/mnras/stv044}, \href
  {http://adsabs.harvard.edu/abs/2015MNRAS.448..568H} {448, 568}

\bibitem[\protect\citeauthoryear{{Hosokawa}, {Omukai}, {Yoshida}  \&
  {Yorke}}{{Hosokawa} et~al.}{2011}]{2011Sci...334.1250H}
{Hosokawa} T.,  {Omukai} K.,  {Yoshida} N.,   {Yorke} H.~W.,  2011, \mn@doi
  [Science] {10.1126/science.1207433}, \href
  {http://adsabs.harvard.edu/abs/2011Sci...334.1250H} {334, 1250}

\bibitem[\protect\citeauthoryear{{Hosokawa}, {Hirano}, {Kuiper}, {Yorke},
  {Omukai}  \& {Yoshida}}{{Hosokawa} et~al.}{2016}]{2016ApJ...824..119H}
{Hosokawa} T.,  {Hirano} S.,  {Kuiper} R.,  {Yorke} H.~W.,  {Omukai} K.,
  {Yoshida} N.,  2016, \mn@doi [\apj] {10.3847/0004-637X/824/2/119}, \href
  {http://adsabs.harvard.edu/abs/2016ApJ...824..119H} {824, 119}

\bibitem[\protect\citeauthoryear{{Hubber} \& {Whitworth}}{{Hubber} \&
  {Whitworth}}{2005}]{2005A&A...437..113H}
{Hubber} D.~A.,  {Whitworth} A.~P.,  2005, \mn@doi [\aap]
  {10.1051/0004-6361:20042428}, \href
  {https://ui.adsabs.harvard.edu/abs/2005A&A...437..113H} {437, 113}

\bibitem[\protect\citeauthoryear{{Hubber}, {Walch}  \& {Whitworth}}{{Hubber}
  et~al.}{2013}]{2013MNRAS.430.3261H}
{Hubber} D.~A.,  {Walch} S.,   {Whitworth} A.~P.,  2013, \mn@doi [\mnras]
  {10.1093/mnras/stt128}, \href
  {http://adsabs.harvard.edu/abs/2013MNRAS.430.3261H} {430, 3261}

\bibitem[\protect\citeauthoryear{{Hummel}, {Stacy}  \& {Bromm}}{{Hummel}
  et~al.}{2016}]{2016MNRAS.460.2432H}
{Hummel} J.~A.,  {Stacy} A.,   {Bromm} V.,  2016, \mn@doi [\mnras]
  {10.1093/mnras/stw1127}, \href
  {http://adsabs.harvard.edu/abs/2016MNRAS.460.2432H} {460, 2432}

\bibitem[\protect\citeauthoryear{{Ishigaki}, {Tominaga}, {Kobayashi}  \&
  {Nomoto}}{{Ishigaki} et~al.}{2018}]{2018ApJ...857...46I}
{Ishigaki} M.~N.,  {Tominaga} N.,  {Kobayashi} C.,   {Nomoto} K.,  2018,
  \mn@doi [\apj] {10.3847/1538-4357/aab3de}, \href
  {https://ui.adsabs.harvard.edu/abs/2018ApJ...857...46I} {857, 46}

\bibitem[\protect\citeauthoryear{{Jappsen}, {Klessen}, {Larson}, {Li}  \& {Mac
  Low}}{{Jappsen} et~al.}{2005}]{2005A&A...435..611J}
{Jappsen} A.-K.,  {Klessen} R.~S.,  {Larson} R.~B.,  {Li} Y.,   {Mac Low}
  M.-M.,  2005, \mn@doi [\aap] {10.1051/0004-6361:20042178}, \href
  {http://adsabs.harvard.edu/abs/2005A%26A...435..611J} {435, 611}

\bibitem[\protect\citeauthoryear{{Johnson} \& {Bromm}}{{Johnson} \&
  {Bromm}}{2006}]{2006MNRAS.366..247J}
{Johnson} J.~L.,  {Bromm} V.,  2006, \mn@doi [\mnras]
  {10.1111/j.1365-2966.2005.09846.x}, \href
  {https://ui.adsabs.harvard.edu/abs/2006MNRAS.366..247J} {366, 247}

\bibitem[\protect\citeauthoryear{{Jones} \& {Bate}}{{Jones} \&
  {Bate}}{2018}]{2018MNRAS.480.2562J}
{Jones} M.~O.,  {Bate} M.~R.,  2018, \mn@doi [\mnras] {10.1093/mnras/sty1969},
  \href {http://adsabs.harvard.edu/abs/2018MNRAS.480.2562J} {480, 2562}

\bibitem[\protect\citeauthoryear{{Kahniashvili}, {Brandenburg}  \&
  {Tevzadze}}{{Kahniashvili} et~al.}{2016}]{2016PhyS...91j4008K}
{Kahniashvili} T.,  {Brandenburg} A.,   {Tevzadze} A. e.~G.,  2016, \mn@doi
  [\physscr] {10.1088/0031-8949/91/10/104008}, \href
  {https://ui.adsabs.harvard.edu/abs/2016PhyS...91j4008K} {91, 104008}

\bibitem[\protect\citeauthoryear{{Karlsson}, {Bromm}  \&
  {Bland-Hawthorn}}{{Karlsson} et~al.}{2013}]{2013RvMP...85..809K}
{Karlsson} T.,  {Bromm} V.,   {Bland-Hawthorn} J.,  2013, \mn@doi [Reviews of
  Modern Physics] {10.1103/RevModPhys.85.809}, \href
  {https://ui.adsabs.harvard.edu/abs/2013RvMP...85..809K} {85, 809}

\bibitem[\protect\citeauthoryear{{Klessen}}{{Klessen}}{2018}]{2018arXiv180706248K}
{Klessen} R.~S.,  2018, arXiv e-prints, \href
  {https://ui.adsabs.harvard.edu/abs/2018arXiv180706248K} {p. arXiv:1807.06248}

\bibitem[\protect\citeauthoryear{{Kouwenhoven}, {Goodwin}, {Parker}, {Davies},
  {Malmberg}  \& {Kroupa}}{{Kouwenhoven} et~al.}{2010}]{2010MNRAS.404.1835K}
{Kouwenhoven} M.~B.~N.,  {Goodwin} S.~P.,  {Parker} R.~J.,  {Davies} M.~B.,
  {Malmberg} D.,   {Kroupa} P.,  2010, \mn@doi [\mnras]
  {10.1111/j.1365-2966.2010.16399.x}, \href
  {https://ui.adsabs.harvard.edu/abs/2010MNRAS.404.1835K} {404, 1835}

\bibitem[\protect\citeauthoryear{{Kratter} \& {Matzner}}{{Kratter} \&
  {Matzner}}{2006}]{2006MNRAS.373.1563K}
{Kratter} K.~M.,  {Matzner} C.~D.,  2006, \mn@doi [\mnras]
  {10.1111/j.1365-2966.2006.11103.x}, \href
  {https://ui.adsabs.harvard.edu/abs/2006MNRAS.373.1563K} {373, 1563}

\bibitem[\protect\citeauthoryear{{Kritsuk}, {Norman}, {Padoan}  \&
  {Wagner}}{{Kritsuk} et~al.}{2007}]{2007ApJ...665..416K}
{Kritsuk} A.~G.,  {Norman} M.~L.,  {Padoan} P.,   {Wagner} R.,  2007, \mn@doi
  [\apj] {10.1086/519443}, \href
  {https://ui.adsabs.harvard.edu/abs/2007ApJ...665..416K} {665, 416}

\bibitem[\protect\citeauthoryear{{Krumholz}}{{Krumholz}}{2014}]{2014MNRAS.437.1662K}
{Krumholz} M.~R.,  2014, \mn@doi [\mnras] {10.1093/mnras/stt2000}, \href
  {http://adsabs.harvard.edu/abs/2014MNRAS.437.1662K} {437, 1662}

\bibitem[\protect\citeauthoryear{{Krumholz}, {McKee}  \& {Klein}}{{Krumholz}
  et~al.}{2004}]{2004ApJ...611..399K}
{Krumholz} M.~R.,  {McKee} C.~F.,   {Klein} R.~I.,  2004, \mn@doi [\apj]
  {10.1086/421935}, \href {http://adsabs.harvard.edu/abs/2004ApJ...611..399K}
  {611, 399}

\bibitem[\protect\citeauthoryear{{Krumholz}, {Klein}  \& {McKee}}{{Krumholz}
  et~al.}{2012}]{2012ApJ...754...71K}
{Krumholz} M.~R.,  {Klein} R.~I.,   {McKee} C.~F.,  2012, \mn@doi [\apj]
  {10.1088/0004-637X/754/1/71}, \href
  {https://ui.adsabs.harvard.edu/abs/2012ApJ...754...71K} {754, 71}

\bibitem[\protect\citeauthoryear{{Larson}}{{Larson}}{1969}]{1969MNRAS.145..271L}
{Larson} R.~B.,  1969, \mn@doi [\mnras] {10.1093/mnras/145.3.271}, \href
  {http://adsabs.harvard.edu/abs/1969MNRAS.145..271L} {145, 271}

\bibitem[\protect\citeauthoryear{{Latif}, {Schleicher}, {Schmidt}  \&
  {Niemeyer}}{{Latif} et~al.}{2013}]{2013MNRAS.432..668L}
{Latif} M.~A.,  {Schleicher} D.~R.~G.,  {Schmidt} W.,   {Niemeyer} J.,  2013,
  \mn@doi [\mnras] {10.1093/mnras/stt503}, \href
  {https://ui.adsabs.harvard.edu/abs/2013MNRAS.432..668L} {432, 668}

\bibitem[\protect\citeauthoryear{{Lepp} \& {Shull}}{{Lepp} \&
  {Shull}}{1984}]{1984ApJ...280..465L}
{Lepp} S.,  {Shull} J.~M.,  1984, \mn@doi [\apj] {10.1086/162013}, \href
  {http://adsabs.harvard.edu/abs/1984ApJ...280..465L} {280, 465}

\bibitem[\protect\citeauthoryear{Lepp, Stancil  \& Dalgarno}{Lepp
  et~al.}{2002}]{Lepp_2002}
Lepp S.,  Stancil P.~C.,   Dalgarno A.,  2002, \mn@doi [Journal of Physics B:
  Atomic, Molecular and Optical Physics] {10.1088/0953-4075/35/10/201}, 35, R57

\bibitem[\protect\citeauthoryear{{Lewis} \& {Bate}}{{Lewis} \&
  {Bate}}{2018}]{2018MNRAS.477.4241L}
{Lewis} B.~T.,  {Bate} M.~R.,  2018, \mn@doi [\mnras] {10.1093/mnras/sty829},
  \href {https://ui.adsabs.harvard.edu/abs/2018MNRAS.477.4241L} {477, 4241}

\bibitem[\protect\citeauthoryear{{Liu} \& {Bromm}}{{Liu} \&
  {Bromm}}{2018}]{2018MNRAS.476.1826L}
{Liu} B.,  {Bromm} V.,  2018, \mn@doi [\mnras] {10.1093/mnras/sty350}, \href
  {https://ui.adsabs.harvard.edu/abs/2018MNRAS.476.1826L} {476, 1826}

\bibitem[\protect\citeauthoryear{{Machida} \& {Doi}}{{Machida} \&
  {Doi}}{2013}]{2013MNRAS.435.3283M}
{Machida} M.~N.,  {Doi} K.,  2013, \mn@doi [\mnras] {10.1093/mnras/stt1524},
  \href {http://adsabs.harvard.edu/abs/2013MNRAS.435.3283M} {435, 3283}

\bibitem[\protect\citeauthoryear{Matthews, Petitpas  \& Aceves}{Matthews
  et~al.}{2011}]{doi:10.1063/1.3628453}
Matthews M.~J.,  Petitpas G.,   Aceves S.~M.,  2011, \mn@doi [Applied Physics
  Letters] {10.1063/1.3628453}, 99, 081906

\bibitem[\protect\citeauthoryear{{McDowell}}{{McDowell}}{1986}]{1986MNRAS.223..763M}
{McDowell} J.~C.,  1986, \mn@doi [\mnras] {10.1093/mnras/223.4.763}, \href
  {http://adsabs.harvard.edu/abs/1986MNRAS.223..763M} {223, 763}

\bibitem[\protect\citeauthoryear{{McKee} \& {Tan}}{{McKee} \&
  {Tan}}{2008}]{2008ApJ...681..771M}
{McKee} C.~F.,  {Tan} J.~C.,  2008, \mn@doi [\apj] {10.1086/587434}, \href
  {http://adsabs.harvard.edu/abs/2008ApJ...681..771M} {681, 771}

\bibitem[\protect\citeauthoryear{{Meru} \& {Bate}}{{Meru} \&
  {Bate}}{2011}]{2011MNRAS.411L...1M}
{Meru} F.,  {Bate} M.~R.,  2011, \mn@doi [\mnras]
  {10.1111/j.1745-3933.2010.00978.x}, \href
  {https://ui.adsabs.harvard.edu/abs/2011MNRAS.411L...1M} {411, L1}

\bibitem[\protect\citeauthoryear{{Nagakura} \& {Omukai}}{{Nagakura} \&
  {Omukai}}{2005}]{2005MNRAS.364.1378N}
{Nagakura} T.,  {Omukai} K.,  2005, \mn@doi [\mnras]
  {10.1111/j.1365-2966.2005.09685.x}, \href
  {https://ui.adsabs.harvard.edu/abs/2005MNRAS.364.1378N} {364, 1378}

\bibitem[\protect\citeauthoryear{{Oesch} et~al.,}{{Oesch}
  et~al.}{2016}]{2016ApJ...819..129O}
{Oesch} P.~A.,  et~al., 2016, \mn@doi [\apj] {10.3847/0004-637X/819/2/129},
  \href {https://ui.adsabs.harvard.edu/abs/2016ApJ...819..129O} {819, 129}

\bibitem[\protect\citeauthoryear{{Omukai}}{{Omukai}}{2001}]{2001ApJ...546..635O}
{Omukai} K.,  2001, \mn@doi [\apj] {10.1086/318296}, \href
  {http://adsabs.harvard.edu/abs/2001ApJ...546..635O} {546, 635}

\bibitem[\protect\citeauthoryear{{Omukai} \& {Nishi}}{{Omukai} \&
  {Nishi}}{1998}]{1998ApJ...508..141O}
{Omukai} K.,  {Nishi} R.,  1998, \mn@doi [\apj] {10.1086/306395}, \href
  {http://adsabs.harvard.edu/abs/1998ApJ...508..141O} {508, 141}

\bibitem[\protect\citeauthoryear{{Omukai}, {Tsuribe}, {Schneider}  \&
  {Ferrara}}{{Omukai} et~al.}{2005}]{2005ApJ...626..627O}
{Omukai} K.,  {Tsuribe} T.,  {Schneider} R.,   {Ferrara} A.,  2005, \mn@doi
  [\apj] {10.1086/429955}, \href
  {http://adsabs.harvard.edu/abs/2005ApJ...626..627O} {626, 627}

\bibitem[\protect\citeauthoryear{{Padoan} \& {Nordlund}}{{Padoan} \&
  {Nordlund}}{2011}]{2011ApJ...730...40P}
{Padoan} P.,  {Nordlund} {\r{A}}.,  2011, \mn@doi [\apj]
  {10.1088/0004-637X/730/1/40}, \href
  {https://ui.adsabs.harvard.edu/abs/2011ApJ...730...40P} {730, 40}

\bibitem[\protect\citeauthoryear{{Peebles}}{{Peebles}}{1968}]{1968ApJ...153....1P}
{Peebles} P.~J.~E.,  1968, \mn@doi [\apj] {10.1086/149628}, \href
  {http://adsabs.harvard.edu/abs/1968ApJ...153....1P} {153, 1}

\bibitem[\protect\citeauthoryear{{Peters}, {Klessen}, {Mac Low}  \&
  {Banerjee}}{{Peters} et~al.}{2010}]{2010ApJ...725..134P}
{Peters} T.,  {Klessen} R.~S.,  {Mac Low} M.-M.,   {Banerjee} R.,  2010,
  \mn@doi [\apj] {10.1088/0004-637X/725/1/134}, \href
  {https://ui.adsabs.harvard.edu/abs/2010ApJ...725..134P} {725, 134}

\bibitem[\protect\citeauthoryear{{Prieto}, {Jimenez}  \& {Mart{\'\i}}}{{Prieto}
  et~al.}{2012}]{2012MNRAS.419.3092P}
{Prieto} J.,  {Jimenez} R.,   {Mart{\'\i}} J.,  2012, \mn@doi [\mnras]
  {10.1111/j.1365-2966.2011.19951.x}, \href
  {https://ui.adsabs.harvard.edu/abs/2012MNRAS.419.3092P} {419, 3092}

\bibitem[\protect\citeauthoryear{Radhakrishnan \& Hindmarsh}{Radhakrishnan \&
  Hindmarsh}{1993}]{osti_15013302}
Radhakrishnan K.,  Hindmarsh A.~C.,  1993, \mn@doi [] {10.2172/15013302}

\bibitem[\protect\citeauthoryear{{Riaz}, {Bovino}, {Vanaverbeke}  \&
  {Schleicher}}{{Riaz} et~al.}{2018}]{2018MNRAS.479..667R}
{Riaz} R.,  {Bovino} S.,  {Vanaverbeke} S.,   {Schleicher} D.~R.~G.,  2018,
  \mn@doi [\mnras] {10.1093/mnras/sty1635}, \href
  {https://ui.adsabs.harvard.edu/abs/2018MNRAS.479..667R} {479, 667}

\bibitem[\protect\citeauthoryear{{Ripamonti} \& {Abel}}{{Ripamonti} \&
  {Abel}}{2004}]{2004MNRAS.348.1019R}
{Ripamonti} E.,  {Abel} T.,  2004, \mn@doi [\mnras]
  {10.1111/j.1365-2966.2004.07422.x}, \href
  {http://adsabs.harvard.edu/abs/2004MNRAS.348.1019R} {348, 1019}

\bibitem[\protect\citeauthoryear{{Saslaw} \& {Zipoy}}{{Saslaw} \&
  {Zipoy}}{1967}]{1967Natur.216..976S}
{Saslaw} W.~C.,  {Zipoy} D.,  1967, \mn@doi [\nat] {10.1038/216976a0}, \href
  {https://ui.adsabs.harvard.edu/abs/1967Natur.216..976S} {216, 976}

\bibitem[\protect\citeauthoryear{{Schauer}, {Whalen}, {Glover}  \&
  {Klessen}}{{Schauer} et~al.}{2015}]{2015MNRAS.454.2441S}
{Schauer} A.~T.~P.,  {Whalen} D.~J.,  {Glover} S.~C.~O.,   {Klessen} R.~S.,
  2015, \mn@doi [\mnras] {10.1093/mnras/stv2117}, \href
  {http://adsabs.harvard.edu/abs/2015MNRAS.454.2441S} {454, 2441}

\bibitem[\protect\citeauthoryear{{Schleicher}, {Banerjee}, {Sur}, {Arshakian},
  {Klessen}, {Beck}  \& {Spaans}}{{Schleicher}
  et~al.}{2010}]{2010A&A...522A.115S}
{Schleicher} D.~R.~G.,  {Banerjee} R.,  {Sur} S.,  {Arshakian} T.~G.,
  {Klessen} R.~S.,  {Beck} R.,   {Spaans} M.,  2010, \mn@doi [\aap]
  {10.1051/0004-6361/201015184}, \href
  {https://ui.adsabs.harvard.edu/abs/2010A%26A...522A.115S} {522, A115}

\bibitem[\protect\citeauthoryear{{Schneider}, {Salvaterra}, {Ferrara}  \&
  {Ciardi}}{{Schneider} et~al.}{2006}]{2006MNRAS.369..825S}
{Schneider} R.,  {Salvaterra} R.,  {Ferrara} A.,   {Ciardi} B.,  2006, \mn@doi
  [\mnras] {10.1111/j.1365-2966.2006.10331.x}, \href
  {https://ui.adsabs.harvard.edu/abs/2006MNRAS.369..825S} {369, 825}

\bibitem[\protect\citeauthoryear{{Schober}, {Schleicher}, {Federrath},
  {Klessen}  \& {Banerjee}}{{Schober} et~al.}{2012a}]{2012PhRvE..85b6303S}
{Schober} J.,  {Schleicher} D.,  {Federrath} C.,  {Klessen} R.,   {Banerjee}
  R.,  2012a, \mn@doi [\pre] {10.1103/PhysRevE.85.026303}, \href
  {https://ui.adsabs.harvard.edu/abs/2012PhRvE..85b6303S} {85, 026303}

\bibitem[\protect\citeauthoryear{{Schober}, {Schleicher}, {Federrath},
  {Glover}, {Klessen}  \& {Banerjee}}{{Schober}
  et~al.}{2012b}]{2012ApJ...754...99S}
{Schober} J.,  {Schleicher} D.,  {Federrath} C.,  {Glover} S.,  {Klessen}
  R.~S.,   {Banerjee} R.,  2012b, \mn@doi [\apj] {10.1088/0004-637X/754/2/99},
  \href {http://adsabs.harvard.edu/abs/2012ApJ...754...99S} {754, 99}

\bibitem[\protect\citeauthoryear{{Silk}}{{Silk}}{1983}]{1983MNRAS.205..705S}
{Silk} J.,  1983, \mn@doi [\mnras] {10.1093/mnras/205.3.705}, \href
  {https://ui.adsabs.harvard.edu/abs/1983MNRAS.205..705S} {205, 705}

\bibitem[\protect\citeauthoryear{{Snaith}, {Park}, {Kim}  \&
  {Rosdahl}}{{Snaith} et~al.}{2018}]{2018MNRAS.477..983S}
{Snaith} O.~N.,  {Park} C.,  {Kim} J.,   {Rosdahl} J.,  2018, \mn@doi [\mnras]
  {10.1093/mnras/sty673}, \href
  {https://ui.adsabs.harvard.edu/abs/2018MNRAS.477..983S} {477, 983}

\bibitem[\protect\citeauthoryear{{Stacy} \& {Bromm}}{{Stacy} \&
  {Bromm}}{2014}]{2014ApJ...785...73S}
{Stacy} A.,  {Bromm} V.,  2014, \mn@doi [\apj] {10.1088/0004-637X/785/1/73},
  \href {http://adsabs.harvard.edu/abs/2014ApJ...785...73S} {785, 73}

\bibitem[\protect\citeauthoryear{{Stacy}, {Greif}  \& {Bromm}}{{Stacy}
  et~al.}{2010}]{2010MNRAS.403...45S}
{Stacy} A.,  {Greif} T.~H.,   {Bromm} V.,  2010, \mn@doi [\mnras]
  {10.1111/j.1365-2966.2009.16113.x}, \href
  {http://adsabs.harvard.edu/abs/2010MNRAS.403...45S} {403, 45}

\bibitem[\protect\citeauthoryear{{Stacy}, {Greif}  \& {Bromm}}{{Stacy}
  et~al.}{2012}]{2012MNRAS.422..290S}
{Stacy} A.,  {Greif} T.~H.,   {Bromm} V.,  2012, \mn@doi [\mnras]
  {10.1111/j.1365-2966.2012.20605.x}, \href
  {http://adsabs.harvard.edu/abs/2012MNRAS.422..290S} {422, 290}

\bibitem[\protect\citeauthoryear{{Stacy}, {Bromm}  \& {Lee}}{{Stacy}
  et~al.}{2016}]{2016MNRAS.462.1307S}
{Stacy} A.,  {Bromm} V.,   {Lee} A.~T.,  2016, \mn@doi [\mnras]
  {10.1093/mnras/stw1728}, \href
  {http://adsabs.harvard.edu/abs/2016MNRAS.462.1307S} {462, 1307}

\bibitem[\protect\citeauthoryear{{Steigman}}{{Steigman}}{2007}]{2007ARNPS..57..463S}
{Steigman} G.,  2007, \mn@doi [Annual Review of Nuclear and Particle Science]
  {10.1146/annurev.nucl.56.080805.140437}, \href
  {https://ui.adsabs.harvard.edu/abs/2007ARNPS..57..463S} {57, 463}

\bibitem[\protect\citeauthoryear{{Sternberg} \& {Neufeld}}{{Sternberg} \&
  {Neufeld}}{1999}]{1999ApJ...516..371S}
{Sternberg} A.,  {Neufeld} D.~A.,  1999, \mn@doi [\apj] {10.1086/307115}, \href
  {https://ui.adsabs.harvard.edu/abs/1999ApJ...516..371S} {516, 371}

\bibitem[\protect\citeauthoryear{{Sur}, {Schleicher}, {Banerjee}, {Federrath}
  \& {Klessen}}{{Sur} et~al.}{2010}]{2010ApJ...721L.134S}
{Sur} S.,  {Schleicher} D.~R.~G.,  {Banerjee} R.,  {Federrath} C.,   {Klessen}
  R.~S.,  2010, \mn@doi [\apjl] {10.1088/2041-8205/721/2/L134}, \href
  {http://adsabs.harvard.edu/abs/2010ApJ...721L.134S} {721, L134}

\bibitem[\protect\citeauthoryear{{Susa}}{{Susa}}{2013}]{2013ApJ...773..185S}
{Susa} H.,  2013, \mn@doi [\apj] {10.1088/0004-637X/773/2/185}, \href
  {http://adsabs.harvard.edu/abs/2013ApJ...773..185S} {773, 185}

\bibitem[\protect\citeauthoryear{{Susa}, {Hasegawa}  \& {Tominaga}}{{Susa}
  et~al.}{2014}]{2014ApJ...792...32S}
{Susa} H.,  {Hasegawa} K.,   {Tominaga} N.,  2014, \mn@doi [\apj]
  {10.1088/0004-637X/792/1/32}, \href
  {http://adsabs.harvard.edu/abs/2014ApJ...792...32S} {792, 32}

\bibitem[\protect\citeauthoryear{{Tomida}, {Tomisaka}, {Matsumoto}, {Hori},
  {Okuzumi}, {Machida}  \& {Saigo}}{{Tomida}
  et~al.}{2013}]{2013ApJ...763....6T}
{Tomida} K.,  {Tomisaka} K.,  {Matsumoto} T.,  {Hori} Y.,  {Okuzumi} S.,
  {Machida} M.~N.,   {Saigo} K.,  2013, \mn@doi [\apj]
  {10.1088/0004-637X/763/1/6}, \href
  {http://adsabs.harvard.edu/abs/2013ApJ...763....6T} {763, 6}

\bibitem[\protect\citeauthoryear{{Truelove}, {Klein}, {McKee}, {Holliman},
  {Howell}  \& {Greenough}}{{Truelove} et~al.}{1997}]{1997ApJ...489L.179T}
{Truelove} J.~K.,  {Klein} R.~I.,  {McKee} C.~F.,  {Holliman} II J.~H.,
  {Howell} L.~H.,   {Greenough} J.~A.,  1997, \mn@doi [\apjl] {10.1086/310975},
  \href {http://adsabs.harvard.edu/abs/1997ApJ...489L.179T} {489, L179}

\bibitem[\protect\citeauthoryear{{Tumlinson}, {Venkatesan}  \&
  {Shull}}{{Tumlinson} et~al.}{2004}]{2004ApJ...612..602T}
{Tumlinson} J.,  {Venkatesan} A.,   {Shull} J.~M.,  2004, \mn@doi [\apj]
  {10.1086/422571}, \href
  {https://ui.adsabs.harvard.edu/abs/2004ApJ...612..602T} {612, 602}

\bibitem[\protect\citeauthoryear{{Turk}, {Smith}, {Oishi}, {Skory}, {Skillman},
  {Abel}  \& {Norman}}{{Turk} et~al.}{2011}]{2011ApJS..192....9T}
{Turk} M.~J.,  {Smith} B.~D.,  {Oishi} J.~S.,  {Skory} S.,  {Skillman} S.~W.,
  {Abel} T.,   {Norman} M.~L.,  2011, \mn@doi [\apjs]
  {10.1088/0067-0049/192/1/9}, \href
  {http://adsabs.harvard.edu/abs/2011ApJS..192....9T} {192, 9}

\bibitem[\protect\citeauthoryear{{Turk}, {Oishi}, {Abel}  \& {Bryan}}{{Turk}
  et~al.}{2012}]{2012ApJ...745..154T}
{Turk} M.~J.,  {Oishi} J.~S.,  {Abel} T.,   {Bryan} G.~L.,  2012, \mn@doi
  [\apj] {10.1088/0004-637X/745/2/154}, \href
  {http://adsabs.harvard.edu/abs/2012ApJ...745..154T} {745, 154}

\bibitem[\protect\citeauthoryear{{Van Borm}, {Bovino}, {Latif}, {Schleicher},
  {Spaans}  \& {Grassi}}{{Van Borm} et~al.}{2014}]{2014A&A...572A..22V}
{Van Borm} C.,  {Bovino} S.,  {Latif} M.~A.,  {Schleicher} D.~R.~G.,  {Spaans}
  M.,   {Grassi} T.,  2014, \mn@doi [\aap] {10.1051/0004-6361/201424658}, \href
  {https://ui.adsabs.harvard.edu/abs/2014A&A...572A..22V} {572, A22}

\bibitem[\protect\citeauthoryear{{Vaytet}, {Tomida}  \& {Chabrier}}{{Vaytet}
  et~al.}{2014}]{2014A&A...563A..85V}
{Vaytet} N.,  {Tomida} K.,   {Chabrier} G.,  2014, \mn@doi [\aap]
  {10.1051/0004-6361/201322855}, \href
  {https://ui.adsabs.harvard.edu/abs/2014A%26A...563A..85V} {563, A85}

\bibitem[\protect\citeauthoryear{{Waagan}}{{Waagan}}{2009}]{2009JCoPh.228.8609W}
{Waagan} K.,  2009, \mn@doi [Journal of Computational Physics]
  {10.1016/j.jcp.2009.08.020}, \href
  {https://ui.adsabs.harvard.edu/abs/2009JCoPh.228.8609W} {228, 8609}

\bibitem[\protect\citeauthoryear{{Waagan}, {Federrath}  \&
  {Klingenberg}}{{Waagan} et~al.}{2011}]{2011JCoPh.230.3331W}
{Waagan} K.,  {Federrath} C.,   {Klingenberg} C.,  2011, \mn@doi [Journal of
  Computational Physics] {10.1016/j.jcp.2011.01.026}, \href
  {http://adsabs.harvard.edu/abs/2011JCoPh.230.3331W} {230, 3331}

\bibitem[\protect\citeauthoryear{{Wang}, {Li}, {Abel}  \& {Nakamura}}{{Wang}
  et~al.}{2010}]{2010ApJ...709...27W}
{Wang} P.,  {Li} Z.-Y.,  {Abel} T.,   {Nakamura} F.,  2010, \mn@doi [\apj]
  {10.1088/0004-637X/709/1/27}, \href
  {http://adsabs.harvard.edu/abs/2010ApJ...709...27W} {709, 27}

\bibitem[\protect\citeauthoryear{{Weidner}, {Kroupa}  \&
  {Maschberger}}{{Weidner} et~al.}{2009}]{2009MNRAS.393..663W}
{Weidner} C.,  {Kroupa} P.,   {Maschberger} T.,  2009, \mn@doi [\mnras]
  {10.1111/j.1365-2966.2008.14258.x}, \href
  {https://ui.adsabs.harvard.edu/abs/2009MNRAS.393..663W} {393, 663}

\bibitem[\protect\citeauthoryear{{W{\"u}nsch}, {Walch}, {Dinnbier}  \&
  {Whitworth}}{{W{\"u}nsch} et~al.}{2018}]{2018MNRAS.475.3393W}
{W{\"u}nsch} R.,  {Walch} S.,  {Dinnbier} F.,   {Whitworth} A.,  2018, \mn@doi
  [\mnras] {10.1093/mnras/sty015}, \href
  {https://ui.adsabs.harvard.edu/abs/2018MNRAS.475.3393W} {475, 3393}

\bibitem[\protect\citeauthoryear{{Yoshida}, {Omukai}, {Hernquist}  \&
  {Abel}}{{Yoshida} et~al.}{2006}]{2006ApJ...652....6Y}
{Yoshida} N.,  {Omukai} K.,  {Hernquist} L.,   {Abel} T.,  2006, \mn@doi [\apj]
  {10.1086/507978}, \href {http://adsabs.harvard.edu/abs/2006ApJ...652....6Y}
  {652, 6}

\bibitem[\protect\citeauthoryear{{Yoshida}, {Oh}, {Kitayama}  \&
  {Hernquist}}{{Yoshida} et~al.}{2007}]{2007ApJ...663..687Y}
{Yoshida} N.,  {Oh} S.~P.,  {Kitayama} T.,   {Hernquist} L.,  2007, \mn@doi
  [\apj] {10.1086/518227}, \href
  {https://ui.adsabs.harvard.edu/abs/2007ApJ...663..687Y} {663, 687}

\makeatother
\end{thebibliography}

    
\appendix
    
\section{Convergence study}
\label{s:append_convergence}
\begin{figure}
\includegraphics[width=1.0\columnwidth]{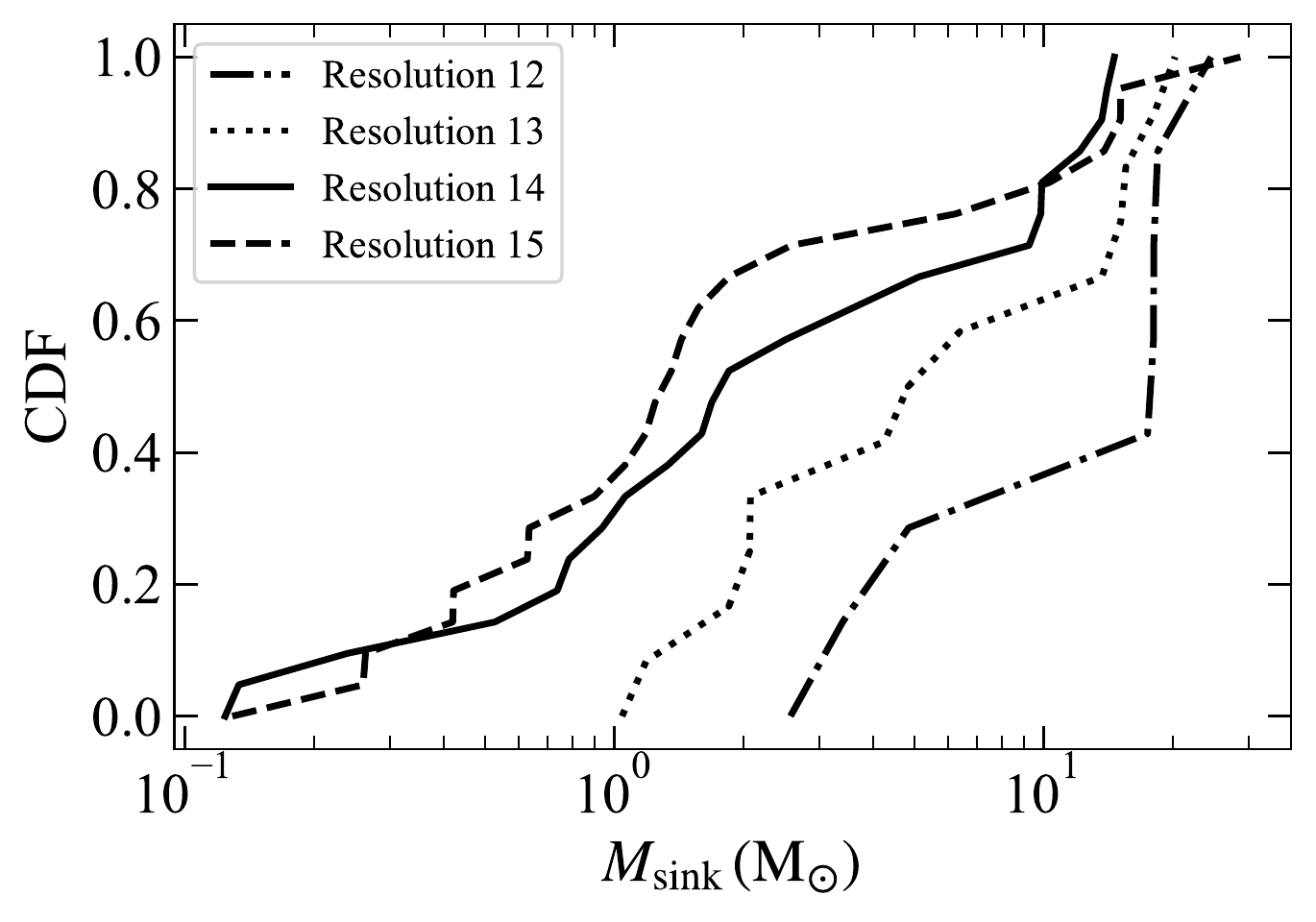}
\caption{CDF of the sink particle mass accumulated over the three runs (A, B, C; see \autoref{tab:converge}) at every resolution.}
\label{fig:convergence}
\end{figure}

It is well known that hydrodynamic simulations of star and galaxy formation can be highly sensitive to the resolution or level of refinement that can be achieved \citep{2008A&A...482..371C,2011MNRAS.411L...1M,2018MNRAS.477..983S}. In the case of primordial star formation, the Jeans scale (and the fragmentation scale) depends on the resolution and thus plays a key role in setting the mass distribution of sink particles \citep{2005SSRv..117..445G,2010MNRAS.403...45S}. Hence, it is necessary to check if the fragmentation we observe in our simulations is scale-dependent. For this purpose, we repeat three runs with variable \gammah (to which we refer in this appendix as runs A, B and C) from our total sample of 40 at four different resolutions with 12, 13, 14 and 15 levels of refinement, respectively (see \autoref{s:hydro} for a description of the levels of refinement). We select these three runs to represent cases of low, medium, and high fragmentation, respectively, at the resolution used in the main text (14 levels). To check for convergence, we compare the state of the runs at $\mbox{SFE}=3.5\%$, rather than 5\% as in the main text. This is a pragmatic choice driven by the high computational cost of attempting to reach $\mbox{SFE}=5\%$ at the highest resolution. \autoref{tab:converge} shows the number of sinks formed, which remains unchanged between resolutions 14 and 15 for all the three runs, suggesting that our results are converged. We find further evidence of convergence at resolution 14 by plotting the CDF of the mass of sink particles accumulated from the three runs at every resolution, as we show in \autoref{fig:convergence}. In fact, the mean sink particle mass also remains the same at resolutions 14 and 15 in all the three runs. While the fragmentation pattern is not identical as we increase the resolution, we do not expect that it should be, since the flows are ultimately chaotic. These differences, however, do not appear to affect the first order characteristics of primordial cloud collapse that we study in this work. 

\begin{table}
\centering
\caption{Summary of outcomes for three sets of variable \gammah runs (A, B, C) carried out at multiple resolutions (12, 13, 14 and 15) with different random turbulent fields. $N_{\mathrm{sink}}$ denotes the number of sink particles at $\mbox{SFE} = 3.5\%$ and $dx$ is the unit cell length at the highest level of refinement corresponding to the resolution used.}
\label{tab:converge}
\begin{tabular}{|lccccr|}
\hline
Property & Resolution & $dx$ & Run A & Run B & Run C\\
\hline
\multirow{5}{*}{$N_{\mathrm{sink}}$} & 12 & $30\,\mathrm{AU}$ & 2 & 4 & 2\\
 & 13 & $15\,\mathrm{AU}$ & 2 & 6 & 5\\
 & 14 & $7.6\,\mathrm{AU}$ & 3 & 6 & 13\\
 & 15 & $3.8\,\mathrm{AU}$ & 3 & 6 & 13\\
\hline
\end{tabular}
\end{table}

\bsp	
\label{lastpage}
\end{document}